\documentclass[12pt,reqno]{amsart}
\usepackage{amsfonts,color}
\usepackage{mathrsfs}
\usepackage{graphicx} %We can use any other package if it is necessary
\usepackage[font=small,labelfont=bf]{caption}
\usepackage{epsfig}
\usepackage{amsmath, amsthm}
\usepackage{amssymb, mathtools}
\usepackage{mathabx}
\usepackage{subfigure,psfrag}
\usepackage[left=2.5cm, right=2.5cm, top=2.5cm, bottom=2.5cm]{geometry}
\numberwithin{equation}{section}
\date{}

\theoremstyle{plain}
\newtheorem{theorem}{Theorem}[section]
\newtheorem{proposition}[theorem]{Proposition}
\newtheorem{lemma}[theorem]{Lemma}
\newtheorem{corollary}[theorem]{Corollary}
\newtheorem{remark}{Remark}[section]
\theoremstyle{definition}
\newtheorem{definition}[theorem]{Definition}
\newtheorem{example}[theorem]{Example}
\numberwithin{equation}{section}

\begin{document}

\def\calL{\mathcal{L}}
\def\calG{\mathcal{G}}
\def\calD{\mathcal{D}}
\def\calJ{\mathcal{J}}
\def\calM{\mathcal{M}}
\def\calN{\mathcal{N}}
\def\calO{\mathcal{O}}
\def\calA{\mathcal{A}}
\def\calS{\mathcal{S}}
\def\calP{\mathcal{P}}
\def\calU{\mathcal{U}}
\def\calK{\mathcal{K}}
\def\frakgl{\mathfrak{gl}}
\def\frako{\mathfrak{o}}
\def\fraku{\mathfrak{u}}
\def\frakg{\mathfrak{g}}
\def\frakso{\mathfrak{so}}
\def\fraksl{\mathfrak{sl}}
\def\fraksp{\mathfrak{sp}}
\def\fraksu{\mathfrak{su}}
\def\F{\mathbb{F}}
\def\R{\mathbb{R}}
\def\N{\mathbb{N}}
\def\C{\mathbb{C}}
\def\M{\mathbb{M}}
\def\H{\mathbb{H}}
\def\P{\mathbb{P}}
\def\al{\alpha}
\def\be{\beta}
\def\p{\partial}
\def\n{\, | \, }
\def\ti{\tilde}
\def\a{\alpha}
\def\r{\rho}
\def\l{\lambda}
\def\hcalG {\hat{\mathcal{G}}}
\def\diag{{\rm diag \/ }}
\def\det{{\rm det \/ }}
\def\sp{{\rm span \/ }}
\def\rd{{\rm d\/}}
\def\K{\nabla}
\def\g{\gamma}
\def\Re{{\rm Re\/}}
\def\a{\alpha}
\def\b{\beta}
\def\d{\delta}
\def\D{\triangle}
\def\e{\epsilon}
\def\g{\gamma}
\def\G{\Gamma}
\def\K{\nabla}
\def\l{\lambda}
\def\L{\Lambda}
\def\n{\,\vert\,}
\def\o{\theta}
\def\w{\omega}
\def\W{\Omega}
\def\ca{{\mathcal{A}}}
\def\cd{{\mathcal{D}}}
\def\cf{{\mathcal{F}}}
\def\cg{{\mathcal{G}}}
\def\ch{{\mathcal{H}}}
\def\ck{{\mathcal{K}}}
\def\cl{{\mathcal{L}}}
\def\cL{{\mathcal{L}}}
\def\cm{{\mathcal{M}}}
\def\cn{{\mathcal{N}}}
\def\co{{\mathcal{O}}}
\def\cp{{\mathcal{P}}}
\def\cs{{\mathcal{S}}}
\def\ct{{\mathcal{T}}}
\def\cu{{\mathcal{U}}}
\def\cv{{\mathcal{V}}}
\def\cx{{\mathcal{X}}}
\def\li{\langle}
\def\ri{\rangle}
\def\n{\ \vert\ }
\def\tr{{\rm tr}}
\def\bs{\bigskip}
\def\ms{\medskip}
\def\ss{\smallskip}
\def\hb{\hfil\break\vskip -12pt}

\def\di{$\diamond$}
\def\ni{\noindent}
\def\ti{\tilde}
\def\p{\partial}
\def\Re{{\rm Re\/}}
\def\Im{{\rm Im\/}}
\def\I{{\rm I\/}}
\def\II{{\rm II\/}}
\def\diag{{\rm diag}}
\def\ad{{\rm ad}}
\def\Ad{{\rm Ad}}
\def\Iso{{\rm Iso}}
\def\Gr{{\rm Gr}}
\def\sgn{{\rm sgn}}

\def\rd{{\rm d\/}}

\def\R{\mathbb{R} }
\def\C{\mathbb{C}}
\def\H{\mathbb{H}}
\def\N{\mathbb{N}}
\def\Z{\mathbb{Z}}
\def\O{\mathbb{O}}
\def\F{\mathbb{F}}

\def\fg{\mathfrak{G}}

\newcommand{\beg}{\begin{example}}
\newcommand{\eeg}{\end{example}}
\newcommand{\bthm}{\begin{theorem}}
\newcommand{\ethm}{\end{theorem}}
\newcommand{\bprop}{\begin{proposition}}
\newcommand{\eprop}{\end{proposition}}
\newcommand{\bcor}{\begin{corollary}}
\newcommand{\ecor}{\end{corollary}}
\newcommand{\blem}{\begin{lemma}}
\newcommand{\elem}{\end{lemma}}
\newcommand{\bca}{\begin{cases}}
\newcommand{\eca}{\end{cases}}
\newcommand{\brem}{\begin{remark}}
\newcommand{\erem}{\end{remark}}
\newcommand{\bpm}{\begin{pmatrix}}
\newcommand{\epm}{\end{pmatrix}}
\newcommand{\bbm}{\begin{bmatrix}}
\newcommand{\ebm}{\end{bmatrix}}
\newcommand{\bvm}{\begin{vmatrix}}
\newcommand{\evm}{\end{vmatrix}}
\newcommand{\bdefn}{\begin{definition}}
\newcommand{\edefn}{\end{definition}}
\newcommand{\bsub}{\begin{subtitle}}
\newcommand{\esub}{\end{subtitle}}
\newcommand{\bex}{\begin{example}}
\newcommand{\eex}{\end{example}}
\newcommand{\ben}{\begin{enumerate}}
\newcommand{\een}{\end{enumerate}}
\newcommand{\bpf}{\begin{proof}}
\newcommand{\epf}{\end{proof}}

\newcommand{\balign}{\begin{align}}
\newcommand{\ealign}{\end{align}}
\newcommand{\baligns}{\begin{align*}}
\newcommand{\ealigns}{\end{align*}}
\newcommand{\beq}{\begin{equation}}
\newcommand{\eeq}{\end{equation}}
\newcommand{\beqs}{\begin{equation*}}
\newcommand{\eeqs}{\end{equation*}}
\newcommand{\beqa}{\begin{eqnarray}}
\newcommand{\eeqa}{\end{eqnarray}}
\newcommand{\beqas}{\begin{eqnarray*}}
\newcommand{\eeqas}{\end{eqnarray*}}

\def\pdo{$\psi$do}

\def\calA{{\mathcal A}}
\def\calB{{\mathcal B}}
\def\calD{{\mathcal D}}
\def\calF{{\mathcal F}}
\def\calG{{\mathcal G}}
\def\calJ{{\mathcal J}}
\def\calK{{\mathcal K}}
\def\calL{{\mathcal L}}
\def\calM{{\mathcal M}}
\def\calN{{\mathcal N}}
\def\calO{{\mathcal O}}
\def\calP{{\mathcal P}}
\def\calR{{\mathcal R}}
\def\calS{{\mathcal S}}
\def\calU{{\mathcal U}}
\def\calV{{\mathcal V}}

\def\li{\langle}
\def\ri{\rangle}

\def\frakP{{\mathfrak{P}}}

\def\half{\frac{1}{2}}
\def\Tr{{\rm Tr\/}}
\def\nkdv{$n\times n$ KdV}

\def \a {\alpha}
\def \b {\beta}
\def \d {\delta}
\def \D {\triangle}
\def \e {\epsilon}
\def \g {\gamma}
\def \G {\Gamma}
\def \K {\nabla}
\def \l {\lambda}
\def \L {\Lambda}
\def \n {\,\vert\,}
\def \N {\,\Vert\,}
\def \o {\theta}
\def\w{\omega}
\def\W{\Omega}
\def \s {\sigma}
\def \S {\Sigma}

\def\ca{{\mathcal {A}}}
\def\cC{{\mathcal {C}}}
\def\cg{{\mathcal {G}}}
\def\ci{{\mathcal {I}}}
\def\ck{{\mathcal {K}}}
\def\cl{{\mathcal {L}}}
\def\cm{{\mathcal {M}}}
\def\cn{{\mathcal {N}}}
\def\co{{\mathcal {O}}}
\def\cp{{\mathcal {P}}}
\def\cs{{\mathcal {S}}}
\def\ct{{\mathcal {T}}}
\def\cu{{\mathcal {U}}}
\def\ch{{\mathcal {H}}}

\def\R{{\mathbb{R}}}
\def\C{{\mathbb{C}}}
\def\H{{\mathbb{H}}}
\def\Z{{\mathbb{Z}}}

\def\Re{{\rm Re\/}}
\def\Im{{\rm Im\/}}
\def\tr{{\rm tr\/}}
\def\Id{{\rm Id\/}}
\def\I{{\rm I\/}}
\def\II{{\rm II\/}}
\def\li{\leftrangle}
\def\ri{rightrangle}
\def\id{{\rm Id}}
\def\gk{\frac{G}{K}}
\def\uk{\frac{U}{K}}

\def\p{\partial}
\def\li{\langle}
\def\ri{\rangle}
\def\ti{\tilde}
\def\i{\/ \rm i }
\def\j{\/ \rm j }
\def\k{\/ \rm k}
\def\n {\ \vert\ }
\def\bu{$\bullet$}
\def\ni{\noindent}
\def\ii{{\rm i\,}}

\def\bs{\bigskip}
\def\ms{\medskip}
\def\ss{\smallskip}

\title[]{Graphic Enumerations and Discrete Painlev\'e Equations via Random Matrix Models}
 
\author{Chuan-Tsung Chan$^\dagger$ \and Hsiao-Fan Liu$^{\ddagger}$}
\address{}
\dedicatory{$^\dagger$ Department of Applied Physics, Tunghai University\\
$^{\ddagger}$Department of Mathematics, National Tsing Hua University\\
$^\dagger$ ctchan@go.thu.edu.tw, $^{\ddagger}$ hfliu@math.nthu.edu.tw}

\date{\today} 

\begin{abstract}

We revisit the enumeration problems of random discrete surfaces (RDS) based on solutions of the discrete equations derived from the matrix models. For RDS made of squares, the recursive coefficients of orthogonal polynomials associated with the quartic matrix model satisfy the discrete type I Painlev\'e equation. Through the use of generating function techniques, we show that the planar contribution to the free energy is controlled by the Catalan numbers. We also develop a new systematic scheme of calculating higher-genus contributions to the topological expansion of the free energy of matrix models. It is important that our exact solutions are valid for finite-$N$ matrix models and no continuous limits are taken within our approach. To show the advantages of our approach, we provide new results of the topological expansion of the free energy for the finite-$N$ cubic matrix model.

\end{abstract}

\maketitle
%%%%%%%%%%            How to add to content line              %%%%%%%%%%%
%\addcontentsline{toc}{chapter}{\protect\numberline{}Appendix}

\lineskip=0.25cm
%%%%%%%%%%%%%%%%%              START HERE        %%%%%%%%%%%%%%%%%%%%%%%%
%%%%%%%%%%%%%%%%%%%%%%%%%%%%%%%%%%%%%%%%%%%%%%%%%%%%%%%%%%%%%%%%%%%%%%%%%
%\large
\section{Introduction and motivation}\

The enumerations of random discrete surfaces (RDS) \cite{Bessis:1980ss,Bessis:1979is} is one of the interesting combinatoric problems, which has important implications to the study of two-dimensional quantum gravity \cite{DiFrancesco:1993cyw}. In the simplest case, we use regular polygons as building blocks to construct closed Riemann surfaces of any topological type (namely, closed surfaces with different genera). From the combinatoric point of view, we are interested in counting the numbers of degeneracy for all possible RDS consisting of $n$ polygons, and of the topological type with genus $h$. For this purpose, it is convenient to define a generating function via a discrete Laplace transformation,
\beq\label{sec1:1}
W(\b,\g):=\sum_{n,h}e^{-n\b+(2-2h)\g}C_{n,h}.
\eeq 
Here $\b$ and $\g$ are dual variables to the total number of polygons, $n$, and the Euler character of the closed surface, $\chi=2-2h$, and $C_{n,h}$ stands for the degeneracy which depends on the type of basic polygons.. 

\begin{figure}[h]
       \begin{center}
       \includegraphics[scale=0.3]{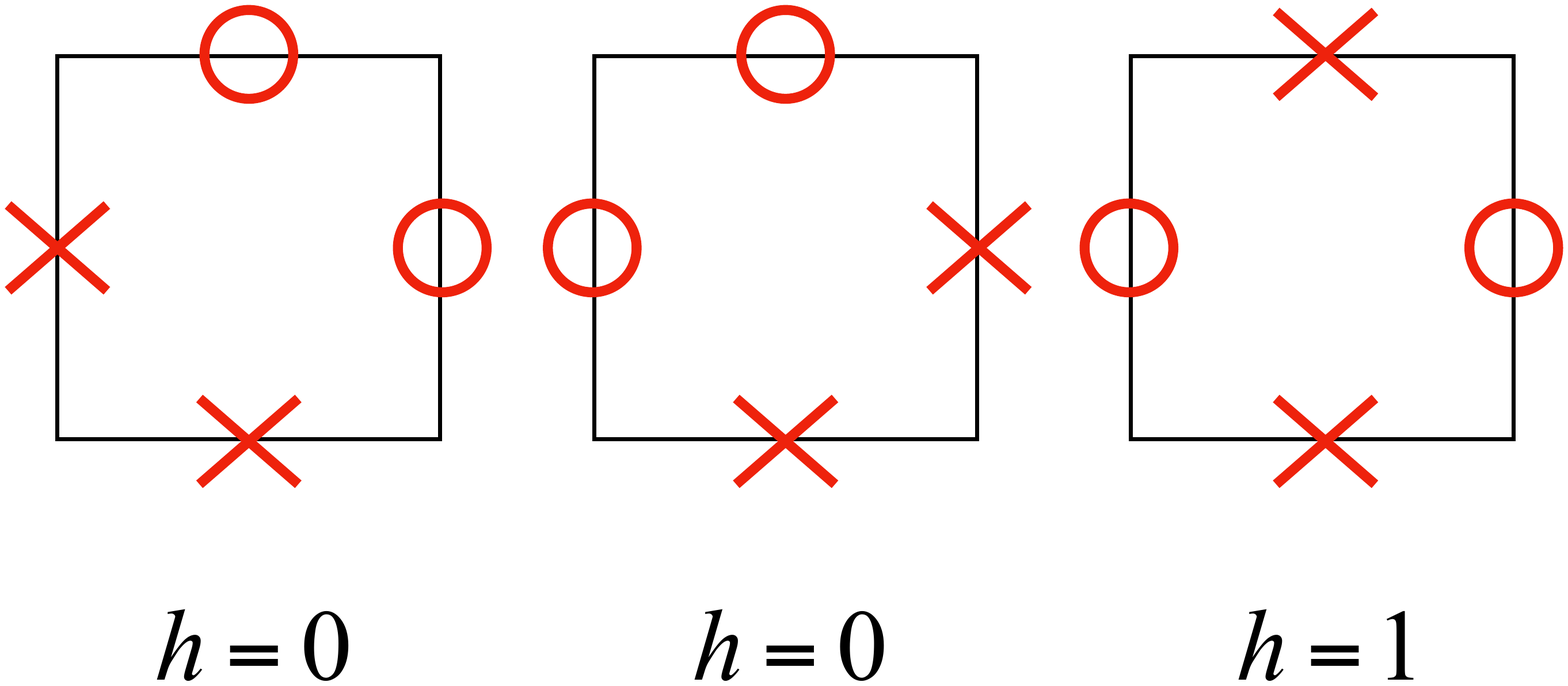}\captionof{figure}{Making closed surfaces by different identifications of the edges of a square}\label{fig1}
       \end{center}
\end{figure}

To illustrate this idea, we use square tiling as an example (Fig.\ref{fig1}). There are three possible ways of identifying (gluing) the four sides of a square ($n=1$) to make a closed surface. Two of them lead to a sphere ($h=0$), and the other gives a torus ($h=1$). 
Consequently, $C_{1,0}=2$ and $C_{1,1}=1$. On the other hand, if we use triangles as basic building blocks for the closed RDS, then it is easy to see that $C_{2k+1,h}=0, \forall h$. Given a fixed number of polygons, it should be easy to imagine that there are only finite possibilities of topological type we can produce. Furthermore, as $n$ grows large, we expect that the degeneracy $C_{n,h}$ will increase fast. Our goal in this paper is to make a quantitive study of these behaviors.   

In fact, the generating function of the enumerations of RDS is closely related to the partition function of the two-dimensional quantum gravity \cite{DiFrancesco:1993cyw,Gross:1989vs,Gross:1989aw,Brezin:1990rb,Douglas:1989ve}. Again, in the simplest case, let us consider the Einstein-Hilbert action in two-dimensional Minkowski space-time:
\beq
\mathcal{S}_{EH} ~ \propto \int (R-2 \Lambda)\sqrt{-g}~ d^2x,
\eeq
where $R$ is the Ricci scalar (twice of the Gaussian curvature) and $\Lambda$ is the cosmological constant. Due to the Gauss-Bonnet theorem \cite{doCarmo:1976}, the integration can be done without knowing an explicit metric of the surface and we get
\beq
\mathcal{S}_{EH}=\g \chi-\b A,
\eeq
where we have absorbed the overall proportional constant into $\g$ (inverse gravitational constant) and $\b$ (cosmological constant).

A popular scenario for quantizing this gravitational theory is to perform functional integrals over all possible metric $g_{\mu\nu}$ (modulus diffeomorphism invariance). Hence we define the partition function,
\beq
W(\b,\g):=\int \frac{\calD g_{\mu\nu}}{[\mbox{diff}]}e^{i\mathcal{S}_{EH}(g_{\mu\nu})} \Rightarrow \sum_{n,h} e^{-n\b+(2-2h)\g}C_{n,h}.
\eeq
In the last step of the equation above, we have performed the Wick rotation ($t\rightarrow i\tau$) to transform the Minkowski space-time into a Euclidean space, and we introduce a discretization procedure to measure the area of a closed surface as the total number of polygons, $n$ \cite{Bessis:1980ss}.

In the pioneer work of Bessis, Itzykson and Zuber \cite{Bessis:1980ss}, the orthogonal polynomial technique of computing random matrix integrals \cite{M60,MG60} was introduced to solve the graphic enumeration problem. The main focus of their work was on the topological expansion of the free energy of the matrix model and they applied continuous approximations in their computations order by order in $1/N^2$. In this paper, however, we would like to propose a different approach based on a direct analysis on the solutions of the recursive coefficients associated with finite-$N$ matrix models. Our approach employs the technique of the generating functions and can be easily adapted to different matrix potentials. In particular, we provide new results on the perturbative series and the topological expansion of the free energy of the cubic matrix model. These correspond to exact (up to genus one) solutions to the graphic enumeration problem for closed discrete surfaces consists of equilateral triangles. 

This paper is organized as follows. We first give a brief introduction to the random matrix model and the discrete difference equations associated with the orthogonal polynomials systems in Sec.$2$. A perturbative analysis of the solutions of the discrete Painlev\'e equation  associated with the recursive coefficients of the of the quartic model, together with the perturbative calculations of the free energy of the quartic model are given in Sec.$3$. Then we proceed with the topological expansion of the free energy of the quartic model in Sec.$4$. Following the same methodology, we study the solutions to the recursive coefficients associated with the orthogonal polynomials, and the perturbative series of the cubic matrix model in Sec.$5$. Sec.$6$ is devoted to the topological expansion of the free energy of the cubic matrix model. We conclude this paper with a brief summary in Sec.$7$. To make the main results of our study clear, we have chosen to collect all technical details in the appendices. In particular, we explain the use of generating functions for the Catalan numbers and its connection with the quartic matrix model in Appendix \ref{app:a}. Similar but more laborious computations for the cubic matrix model is presented in Appendix \ref{app:b}.

\section{Graphic enumerations, random matrix models and discrete Painlev\'e equations}\

\subsection{Wick theorem for random matrix integrals}\label{subsec1}\

\vspace{0.5cm}
In order to compute the generating function Eq.\eqref{sec1:1} for the graphical enumeration problem, we resort to the quantum field theory techniques developed  in \cite{Bessis:1980ss,tHooft:1973alw,Brezin:1977sv}. The basic idea is to map a random matrix integral into a counting device for discrete surfaces consisting of regular polygons. To see this, consider the partition function of the quartic Hermitian matrix model,
\beq\label{sec2:1b}
Z(g,N):=\int \calD M e^{-N\tr\left(\frac{M^2}{2}-\frac{g}{4}M^4\right)},
\eeq
where $M$ is a $N \times N$ Hermitian matrix and $V(x):=\frac{x^2}{2}-\frac{g}{4}x^4$ is the quartic potential with coupling constant $g$. We choose the Cartesian integration measure for the matrix integral as,
\beq
\calD M:=\left(\prod_{k=1}^N dm_{kk}\right)\left[\prod_{j<i}\left(d~\Re m_{ij}\right)\left(d~\Im m_{ij}\right)\right],
\eeq
which is invariant under the unitary conjugation (gauge transformation),
\beq
M \rightarrow M^U:=U^\dag MU \Rightarrow \calD M^U=\calD M.
\eeq

The partition function, treated as a formal power series in $g$, can be computed via the standard perturbative approach,
\beqa
\frac{Z(g,N)}{Z(0,N)}
&=&\frac{1}{Z(0,N)}\sum_{k=0}^\infty \left(\frac{gN}{4}\right)^k \frac{1}{k!} \int \calD M\left(\tr M^4\right)^k e^{-\frac{N}{2}\tr (M^2)}\nonumber\\
&=:&1+z_1(N)\left(\frac{g}{4}\right)+z_2(N)\left(\frac{g}{4}\right)^2+O(g^3),
\eeqa
and the generalized moments, $z_1(N), z_2(N), \cdots$, can be computed via the Wick theorem \cite{Bessis:1980ss,tHooft:1973alw,Brezin:1977sv}.

For this purpose, we first define the propagator as the second moment of the matrix integral over Gaussian weight,
\beq\label{sec2:1a}
<m_{ij}m_{kl}>:=\frac{1}{Z(0,N)} \int \calD M m_{ij}m_{kl} e^{-\frac{N}{2}\tr (M^2)}=\frac{1}{N}\d_{il}\d_{jk},
\eeq
and
\beqa
z_1(N)&=&N\sum_{i,j,k,l}<m_{ij}m_{jk}m_{kl}m_{li}>\nonumber\\
&=&N\sum_{i,j,k,l} \left[\undergroup{m_{ij}m_{jk}}\overgroup{m_{kl}m_{li}}+\lefteqn{\undergroup{\phantom{m_{ij}m_{jk}m_{kl}}}}m_{ij}\overgroup{m_{jk}m_{kl}m_{li}}+\undergroup{m_{ij}\overgroup{m_{jk}m_{kl}}m_{li}}\right]\nonumber\\
&=&\frac{N}{N^2}\sum_{i,j,k,l}\left(\d_{ik}\d_{jj}\d_{ki}\d_{ll}+\d_{il}\d_{jk}\d_{ji}\d_{kl}+\d_{ii}\d_{jl}\d_{jl}\d_{kk}\right)\nonumber\\
&=&2N^2+1.
\eeqa

Similarly, the second-order contribution to the partition function, $z_2(N)$, is given by contractions among eight matrix elements and it is given as 
\beqa\label{sec2:1}
z_2(N)&=&\frac{N^2}{2!}\sum_{i,j,k,l, p,q,r,s} <m_{ij}m_{jk}m_{kl}m_{li}m_{pq}m_{qr}m_{rs}m_{sp}>\nonumber\\
&=&2N^4+20N^2+\frac{61}{2}.
\eeqa

Note that the leading term, $2N^4$, in Eq.\eqref{sec2:1}, is given by the product of the first-order leading term
\beq
<m^8> \sim <m^4><m^4> \sim \frac{4N^6}{N^4}.
\eeq
The free energy of the quartic matrix model, defined as the logarithm of the normalized partition function, counting only closed connected discrete surfaces \cite{Bessis:1980ss,tHooft:1973alw,Brezin:1977sv} made from squares, is given as
\beqa
F(g,N)&:=&\ln\left[\frac{Z(g,N)}{Z(0,N)}\right]\nonumber\\
&=&(2N^2+1)\left(\frac{g}{4}\right)+(18N^2+30)\left(\frac{g}{4}\right)^2+O\left(\frac{g^3}{4^3}\right).
\eeqa

\subsection{Diagrammatic expansion of the partition function of the matrix model}\
\vspace{0.5cm}   

 The reason why the partition function of a random matrix model lead to solution of the graphical enumeration problem can be seen from the diagrammatic expansion of the matrix integrals. 
 
\begin{figure}[h]
       \begin{center}
       \includegraphics[scale=0.3]{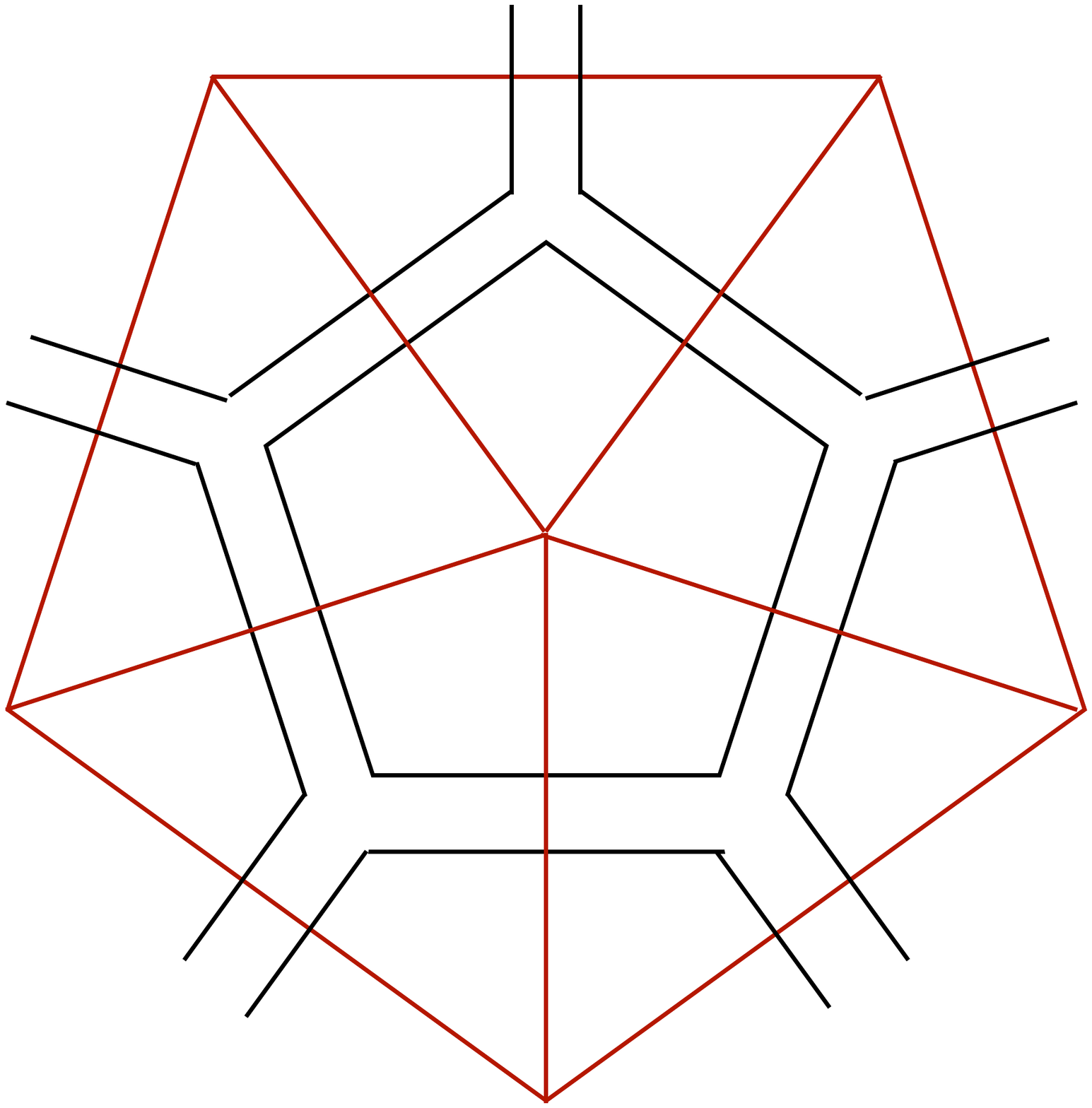}
       \captionof{figure}{Dual transformation from the fat-graph (black) to a RDS (red)}\label{fig:1}
       \end{center}
\end{figure}

 First of all, for closed discrete surface made of identical regular $m$ polygons, we use a simple matrix potential, $V(x)=\frac{x^2}{2}-\frac{g}{m}x^m$, and the weight function, $e^{-NV(x)}$, can be expanded as a formal  series in $g$,
\beq
e^{-N(\frac{x^2}{2}-\frac{g}{m}x^m)}=\sum_{k=0}^\infty \left(\frac{gN}{m}\right)^k  \frac{\left(x^m\right)^k}{k!} e^{-\frac{N}{2}x^2}.
\eeq 
A standard practice in quantum field theory is then using the Gaussian part to construct propagator and treating $\frac{g}{m}x^m$ as an interaction vertex. The Wick theorem, discussed in the previous section, implies joining various vertices by suitable number of propagators. The random matrix integrals differs from the path integrals of a point particle in that the matrix propagator carries two sets of indices Eq.\eqref{sec2:1a}, and the point-particle Feynman diagrams become a fat-diagram consisting of irregular polygons, as shown in Fig. \ref{fig:1} with black lines. 

At the last step, we need to map the fat-diagram into a random discrete surfaces by a dual transformation. That is, for each irregular polygon in the fat-diagram, we assign the center of it as one of the vertices of the RDS. By connecting these RDS vertices with edges, we see that each matrix vertex is enclosed by a $m$-polygon, and the fat-diagram, after this dual transformation, becomes a random discrete surface (the red lines in Fig.\ref{fig:1}). 
In summary, the dual transformation trade the vertices in the fat graph into regular $m$-polygons (faces), the irregular polygon in the fat graph into a vertex, and the propagator in the flat graph will intersect with an unique edge of the RDS.  
 
From these correspondences, we can see that, according to the standard Feynman rule for the random matrix integrals:
\begin{itemize}
\item Each vertex of a fat graph contributes a factor of $gN$ (note our normalization of the coupling in Eq.\eqref{sec2:1b}),
\item Each propagator of a fat graph contributes a factor of $1/N$,
\item Each internal loop demands a sum over dummy matrix indices and contributes a factor of $N$, and the total number of the internal loops within a given fat diagram is equal to the total number of the vertices of the RDS. 
\end{itemize}
Putting all these ingredients together, we find that for each fat diagram, we get 
\beq
(gN)^F(N^{-1})^E(N)^V=g^FN^{V-E+F}=g^nN^\chi,
\eeq
where we have used the Euler character formula of a closed surface, 
\beq
\chi=V-E+F,
\eeq  
and $n$ denotes the total number of the faces (area) of the RDS.
 
The diagrammatic expansion of the partition function of a random matrix integral generally includes disconnected diagrams, which are unions of closed RDS's. In order to remove these repetitions, we define the free energy as the logarithm of the normalized partition function,
\beq
F(g,N):=\ln\left[\frac{Z(g,N)}{Z(0,N)}\right].
\eeq  
It is an established result that $F(g,N)$ counts only connected RDS \cite{Bessis:1980ss,tHooft:1973alw,Brezin:1977sv}. Thus, we show that the diagrammatic expansion of the free energy of a random matrix integral, viewed as a function of the coupling constant, $g$, and the size of the matrix, $N$, can be expressed as a formal series
\beq
F(g,N)=\sum_{n,h} g^nN^{2-2h}C_{n,h}.
\eeq 
Comparing with the generating function of the graphic enumeration, Eq.\eqref{sec1:1}, we found an identification between matrix model parameters and the two-dimensional quantum gravity data as follows:
\beq
g=e^{-\b}, \quad N=e^\g, 
\eeq 
and $C_{n,h}$ is precisely the degeneracy number we mentioned in the first section.

\subsection{Orthogonal polynomial technique for random matrix integrals}\
\vspace{0.5cm}

As shown in Sec.\ref{subsec1}, the perturbative series of the random matrix integrals can be evaluated with the Wick contractions and in principle manageable via computer program. Nevertheless, the apparent complications associated with the fast growing of number of contractions and the need of some partial re-summation (e.g. topological expansion in 1/$N^2$) of the free energy prompts us for other resolution.

It turns out, at least for the case of the Hermitian matrix model, it is possible to obtain an exact expression of the partition functions for the random matrix integrals by means of the orthogonal polynomial techniques \cite{M60,MG60}. The essential idea is that by a change of variables from the $N^2$ matrix elements, $m_{ij}$, to the eigenvalues, $\l_k$, and the associated eigenvectors, $\vec{v}_k$, $1\leq k\leq N$, in the random matrix integral \eqref{sec_2:1}, the associated Jacobian, namely, the square of the Vandermonde determinant, can be expressed as a joint product of the orthogonal pairing (inner product) of the monic orthogonal polynomials defined as 
\beq
\int P_m(x)P_n(x)e^{-NV(x)}~dx=h_n\d_{mn}.
\eeq
\beqa
\calD M&=&\frac{1}{N!}\left(\calD U\right)\left(\prod_{k=1}^N d \l_k\right)\left(\bigtriangleup(\l)\right)^2, \label{sec_2:1}\\
\D(\l)&:=&\det \left(\begin{array}{cccc}
                     1 & \l_1^{0} & \cdots & \l_{N-1}^{0} \\
                     1 & \l_1^{1} & \cdots & \l_{N-1}^{1} \\
                     \vdots & \vdots &  & \vdots \\
                     1 & \l_1^{N-1} & \cdots & \l_{N-1}^{N-1}
                     \end{array}\right)\\
        &=&\det \left(\begin{array}{cccc}
                     1 & P_0(\l_1) & \cdots & P_0(\l_{N-1}) \\
                     1 & P_1(\l_1) & \cdots & P_1(\l_{N-1}) \\
                     \vdots & \vdots &  & \vdots \\
                     1 & P_{N-1}(\l_1) & \cdots & P_{N-1}(\l_{N-1})
                     \end{array}\right).
\eeqa

\beqa
&&\int \calD M e^{-N \tr V(M)}\nonumber\\
&=&\frac{1}{N!}\int \prod_{k=1}^N d \l_k\left[\det \left(\begin{array}{cccc}
                     1 & P_0(\l_1) & \cdots & P_0(\l_{N-1}) \\
                     1 & P_1(\l_1) & \cdots & P_1(\l_{N-1}) \\
                     \vdots & \vdots &  & \vdots \\
                     1 & P_{N-1}(\l_1) & \cdots & P_{N-1}(\l_{N-1})
                     \end{array}\right)\right]^2e^{-N\left[\sum_{l=1}^N V(\l_l)\right]}\nonumber\\
&=&\frac{1}{N!}\left(\prod_{k=0}^{N-1} h_k\right).
\eeqa
Thus, if we can compute the normalization constants of the monic orthogonal polynomials associated with a given matrix potential, $V(x)$, the free energy of the corresponding random matrix integral is given by 
\beq
F(g,N):=\ln\left[\frac{Z(g,N)}{Z(0,N)}\right]=\sum_{k=0}^{N-1}\ln\left[\frac{h_k(g,N)}{h_k(0,N)}\right].
\eeq

At this stage, we have transformed the original graphic enumeration problem to the problem of computing normalization constants of the associated orthogonal polynomials. In the next step, we show how to calculate the normalization constants from the recursive coefficients of the three-term relation.

\subsection{Normalization constants and recursive coefficients of the monic orthogonal polynomial}\
\vspace{0.5cm}

It is a well-known fact that any orthogonal polynomial system satisfies a three-term recursive relation,
\beq\label{sec_2:2}
xP_n(x)=P_{n+1}(x)+\b_nP_n(x)+\a_nP_{n-1}(x),
\eeq
where $\a_n,\b_n$ will be referred to as recursive coefficients and we choose to work with monic orthogonal polynomials
\beq
P_n(x)=x^n+\cdots.
\eeq
The recursive coefficient, $\a_n$, is related to the normalization constant $h_n$ by 
\beqa
h_n&=& \int P_n(x)\left[xP_{n-1}(x)\right]e^{-NV(x)}~dx\nonumber\\
&=&\int \left[\a_nP_{n-1}(x)\right]P_{n-1}(x)e^{-NV(x)}~dx\nonumber\\
&=&\a_n h_{n-1},\\
 \Rightarrow \a_n&=&\frac{h_n}{h_{n-1}}.\label{sec_2:3}
\eeqa

On the other hand, both recursive coefficients, $\a_n,\b_n$ satisfy a coupled set of difference equations. Taking cubic matrix model $V(x)=\frac{x^2}{2}-\frac{g}{3}x^3$ as an example, we have 

\bthm
The recursive coefficients, $\a_n,\b_n$, of the monic orthogonal polynomials 
associated with the cubic matrix model, Eq.\eqref{sec_2:2}, satisfy the {\rm d-PII} equations,
\beqa
\frac{n}{N}&=&\a_n\left[1-g(\b_n+\b_{n-1})\right],\label{sec_2:4}\\
\b_n&=&g(\a_{n+1}+\b_n^2+\a_n).\label{sec_2:5}
\eeqa
\ethm
\bpf
To derive an equation relating $\b_n$ and $\a_n$, we compute the projection of $\frac{d}{dx}P_n(x)$ on $P_{n-1}(x)$,
\beqa
nh_{n-1}
&=&\int P_{n-1}(x)\left[\frac{d}{dx}P_n(x)\right]e^{-NV(x)}~dx\nonumber\\
&=&-\int \left[\frac{d}{dx}P_{n-1}(x)\right]P_n(x)e^{-NV(x)}~dx+N\int P_{n-1}(x)P_n(x)\left(\frac{dV}{dx}\right)e^{-NV(x)}~dx\nonumber\\
&=&N\left[h_n-gh_n(\b_{n-1}+\b_n)\right].
\eeqa
Divide both sides by $h_{n-1}$, we get Eq.\eqref{sec_2:4}.
We then compute the projection of $\frac{d}{dx}P_n(x)$ on $P_n$: 
\beqa
0&=&\int P_n(x)\left[\frac{d}{dx}P_n(x)\right] e^{-NV(x)}~dx\nonumber\\
&=&-\int \left[\frac{d}{dx}P_n(x)\right]P_n(x) e^{-NV(x)}~dx+N\int P_n(x)P_n(x)V'(x)e^{-NV(x)}~dx\nonumber\\
&=&N\int P_n(x)P_n(x)(x-gx^2)e^{-NV(x)}~dx\nonumber\\
&=&N\left[\b_n h_n-g\left(h_{n+1}+\b_n^2 h_n+\a_n^2h_{n-1}\right)\right]\\
\Rightarrow \b_n&=&g(\a_{n+1}+\b_n^2+\a_n).\nonumber
\eeqa
\epf

\bthm
The coupled difference equations Eqs.\eqref{sec_2:4}, \eqref{sec_2:5} are equivalent to the discrete type II Painlev\'e {\rm (d-PII)} equations by a change of variables.
\beqas
p_n&:=&\a_n, \quad q_n:=\frac{\b_{n-1}-\sqrt{t}}{\sqrt{2}i},\\
t&:=&4g^2, \quad a(n):=\left(\frac{2\sqrt{2}  g}{i N}\right)n.
\eeqas
\ethm
\bpf
By substituting the definitions of the variables $(\{\a_n,\b_n\} \rightarrow \{q_n,p_n\})$ and the parameters $(\{g,N\} \rightarrow \{t,a(n)\})$, we get 
\beq
p_n(q_n+q_{n+1})=-a(n), \mbox{ and } p_{n+1}+p_n=2q_{n+1}^2+t.
\eeq
Indeed, these new coupled difference equations are compatibility conditions of the Lax pairs 
associated with the d-PII (d-P($A_1^{(1)}/E_7^{(1)}$)) Hamiltonian 
\beq
H=\frac{p^2}{2}-\left(q^2+\frac{t}{2}\right)p-aq
\eeq
in \cite{KNY2017}.
\epf

\begin{figure}[h]
       \begin{center}
       \includegraphics[scale=0.4]{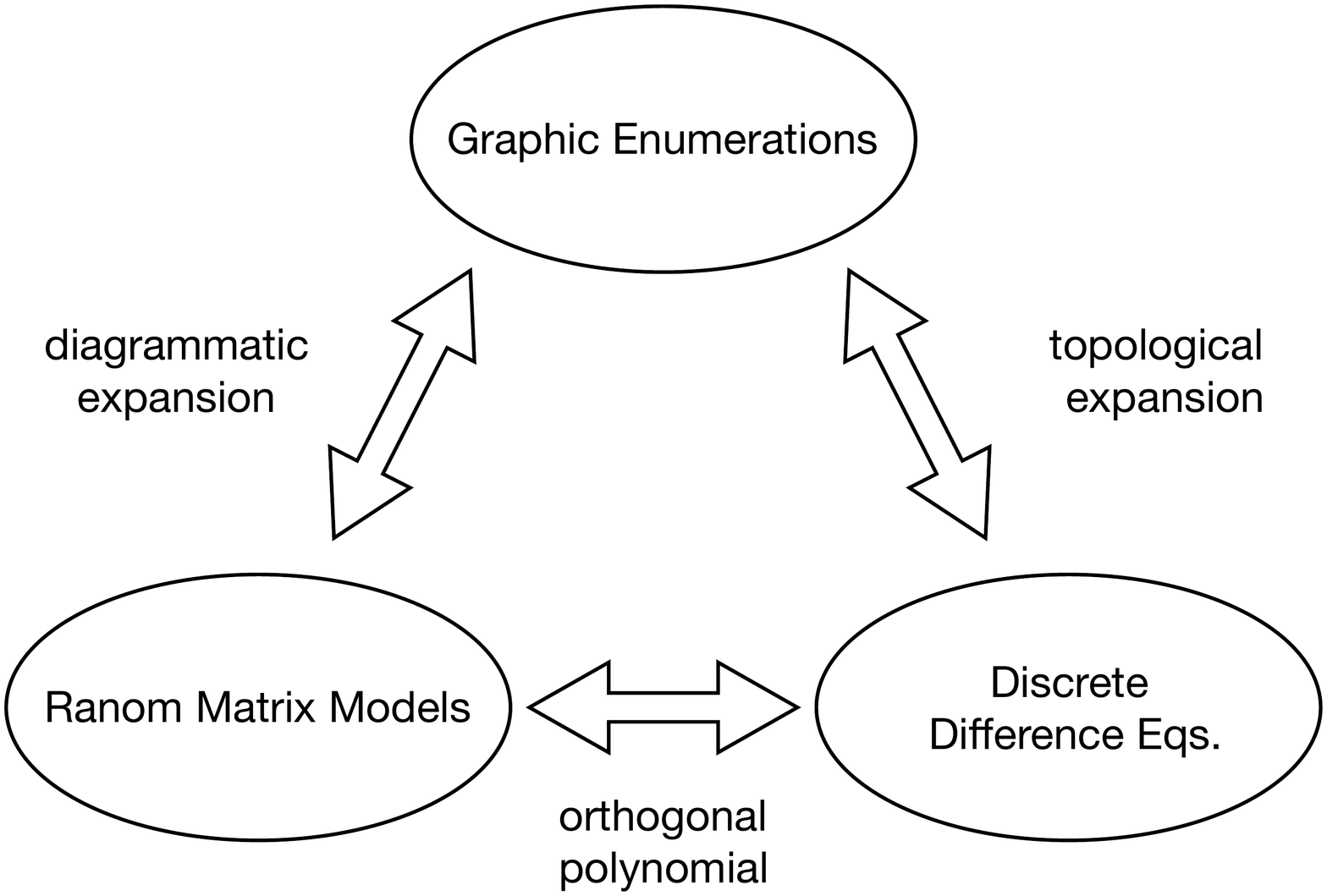}
       \captionof{figure}{Interconnections between three subjects under our study}\label{fig:2}
       \end{center}
\end{figure}

In the case of the quartic matrix model, $V(x)=\frac{x^2}{2}-\frac{g}{4}x^4$, there is only one type of recursive coefficient due to parity conservation,
\beq
xP_n(x)=P_{n+1}(x)+\g_n P_{n-1}(x).
\eeq
In addition to the same normalization constant relation, Eq.\eqref{sec_2:3},
\beq
\g_n=\frac{h_n}{h_{n-1}},
\eeq
the recursive coefficient, $\g_n$, satisfies the celebrated discrete type I Painlev\'e (d-PI) equation.
\bthm\cite{Fokas:1990wb}
The recursive coefficients of the monic orthogonal polynomials associated with 
the quartic matrix model, satisfies the {\rm d-PI} equation, 
\beq\label{sec2-4:1}
\frac{n}{N}=\g_n-g \g_n(\g_{n+1}+\g_n+\g_{n-1}).
\eeq
\ethm
Having explained all basic ingredients of this work, we use the following diagram (Fig.\ref{fig:2}) to summarize the interconnections among various subjects.

Starting from next section, we shall perform both perturbative (in $g$) and topological (in $1/N^2$) studies of the solutions of the recursive coefficients, $\g_n$ for the quartic matrix model, and $\a_n,\b_n$ for the cubic matrix model. Once these answers are calculated, we shall obtain the free energy of the random matrix model, and hence the generating function of the graphic enumeration problem as
\beq
W(\b,\g)=F(e^{-\b},e^{\g}),
\eeq 
and
\beq\label{sec2:2b}
F(g,N)=N \ln\left[\frac{h_0(g,N)}{h_0(0,N)}\right]+\sum_{k=1}^{N-1}\left(N-k\right) \ln \left[\frac{\g_k(g,N)}{\g_k(0,N)}\right].
\eeq

%\section{}

%\input{sec_3.tex}

\section{Perturbative expansion of the free energy of the quartic model}\

In the previous sections, we have established a connection between the free energies of the matrix models and the recursive coefficients of the associated monic orthogonal polynomial systems. Here we shall perform a perturbative analysis of the discrete difference equation (d-PI) to obtain a series solution of the recursive coefficients of the quartic matrix model.

To begin with, the d-PI equation is written as 
\beq
\frac{n}{N}=\g_n-g\g_n(\g_{n+1}+\g_n+\g_{n-1}).
\eeq
If we assume $g$ is a small umber and treat the recursive coefficient $\g_n$ as a power series in $g$,
\beq\label{sec3:1}
\g_n(g,N)=\sum_{k=0}^\infty \g_n^{(k)}(N) g^k,
\eeq
then the d-PI equation can be expressed as a infinite set of coupled difference equation.
\beqa
\frac{n}{N}&=&\g_n^{(0)},\\
\g_n^{(1)}&=&\g_n^{(0)}(\g_{n+1}^{(0)}+\g_n^{(0)}+\g_{n-1}^{(0)}),\\
\g_n^{(2)}&=&\g_n^{(0)}(\g_{n+1}^{(1)}+\g_n^{(1)}+\g_{n-1}^{(1)})+\g_n^{(1)}(\g_{n+1}^{(0)}+\g_n^{(0)}+\g_{n-1}^{(0)}),\\
&\vdots&\nonumber\\
\g_n^{(k)}&=&\sum_{m=0}^{k-1}\g_n^{(m)}\left[\g_{n+1}^{(k-m-1)}+\g_n^{(k-m-1)}+\g_{n-1}^{(k-m-1)}\right].\label{sec3:2a}
\eeqa
By iteration, we get the solutions of the perturbative series as follows
\beqa
\g_n^{(0)}&=&\frac{n}{N},\quad \g_n^{(1)}=\frac{3n^2}{N^2}, \quad \g_n^{(2)}=\frac{18n^3+6n}{N^3},\nonumber\\
\g_n^{(3)}&=&\frac{135n^4+162n^2}{N^4},\quad \g_n^{(4)}=\frac{18(63n^5+174n^3+35n)}{N^5}, \quad \cdots.
\eeqa
From these explicit results, together with the deductive reasoning, we find that 
\beq
\g_n(g,N)=\sum_{k=0}^\infty \g_n^{(k)}(N)g^k=\sum_{k=0}^\infty \left(\frac{\mbox{polynomial in }n \mbox{ with } \deg k+1}{N}\right)\left(\frac{g}{N}\right)^k.
\eeq 
This fact will be useful when we study the topological expansions of the matrix model free energies later in Sec.$4$.

To compute the free energy, we also need the normalization constant for the zeroth orthogonal polynomial, which can be expressed as a formal power series in $g$,
\beqa
h_0(g,N)&:=&\int_{-\infty}^\infty e^{-NV(x)}~dx \Rightarrow \int_{-\infty}^\infty e^{-N(\frac{x^2}{2}-\frac{g}{4}x^4)}~dx\nonumber\\
&=&\sqrt{\frac{2\pi}{N}}\left[1+\left(\frac{3}{4}\right)\frac{g}{N}+\left(\frac{105}{32}\right)\frac{g^2}{N^2}+\left(\frac{105\cdot33}{2^7}\right)\frac{g^3}{N^3}+O\left(\frac{g^4}{N^4}\right)\right].
\eeqa
This implies
\beqa
&&\ln h_0(g,N)-\ln h_0(0,N)\nonumber\\
&=&3\left(\frac{g}{4N}\right)+48\left(\frac{g}{4N}\right)^2+1584\left(\frac{g}{4N}\right)^3+O\left(\frac{g^4}{4^4N^4}\right).
\eeqa
The logarithms of the recursive coefficients are
\beqa
&&\ln \g_n(g,N)-\ln \g_n(0,N)\\
&=&\ln\left[1+\frac{\g_n^{(1)} }{\g_n^{(0)} }g+\frac{\g_n^{(2)} }{\g_n^{(0)} }g^2+\frac{\g_n^{(3)} }{\g_n^{(0)} }g^3+O(g^4)\right]\\
&=&\left(\frac{3n}{N}\right)g+\left(\frac{27n^2+12}{2N^2}\right)g^2+\left(\frac{90n^3+144n}{N^3}\right)g^3+O(g^4).
\eeqa
Putting all ingredients together, and recalling a master formula for the sum of powers of integers \cite{BA96}
\beq\label{sec3:1a}
1^k+2^k+\cdots+(N-1)^k=\frac{N^{k+1}}{k+1}-\frac{N^k}{2}+\frac{kN^{k-1}}{12}+O(N^{k-3}), \quad k \geq 3,
\eeq
we derive the first three terms of perturbative series (in $g$) of the free energy in the case of quartic model,
\beqa
F(g,N)&=&N\ln \left[\frac{h_0(g,N)}{h_0(0,N)}\right]+\sum_{n=1}^{N-1} (N-n)\ln \left[\frac{\g_n(g,N)}{\g_n(0,N)}\right]\\
&=&(2N^2+1)\left(\frac{g}{4}\right)+(18N^2+30)\left(\frac{g}{4}\right)^2+(288N^2+1056+\frac{240}{N^2})\left(\frac{g}{4}\right)^3+O(g^4).\nonumber
\eeqa

\section{Topological expansion of the free energy of the quartic model}\

In order to derive a topological (1/$N^2$) expansion of the free energy of the quartic model, we first examine the structure of the discrete Painlev\'e equation, Eq.\eqref{sec2-4:1}, of the recursive coefficients. By a rescaling
\beq
\ti\g_n := g \g_n \Leftrightarrow \g_n=\frac{1}{g} \ti\g_n, 
\eeq
the d-PI equation becomes 
\beq
n \e =\ti\g_n-\ti\g_n(\ti\g_{n+1}+\ti\g_n+\ti\g_{n-1}), \quad \e:=\frac{g}{N}.
\eeq
Consequently, the recursive coefficients can be expressed as a Taylor series of $\frac{g}{N}$,
\beq
\g_n%=\sum_{m=0}^\infty\left[n^{m+1}3^mC_m^{(0)+n^{m-1}3^mC_m^{(1)}}\right]\frac{g^m}{N^{m+1}}
=\frac{1}{g}\sum_{m=0}^\infty \g_{n,m}\e^{m+1}.
\eeq
Note that $\g_{n,m}=\g_n^{(m)}N^{m+1}$ in Eq.\eqref{sec3:1}. Substituting these Taylor series back to the d-PI equation, Eq.\eqref{sec3:2a}, we get 
\beq\label{sec_5:1}
\g_{n,m+1}=\sum_{l=0}^m \g_{n,m-l}(\g_{n-1,l}+\g_{n,l}+\g_{n+1,l}), \quad m\geq 0.
\eeq
%By straightforward iteration, we get 
%\beq
%\g_{n,0}=n, \quad \g_{n,1}=, \quad \g_{n,2}=.
%\eeq

In general, by mathematical induction, we can show that $\g_{n,m}$ are polynomials in $n$ of degree $m+1$ with definite parity,
\beq
\g_{-n,m}=(-1)^{m+1}\g_{n,m}.
\eeq
Thus, we can decompose the Taylor coefficients of the recursive coefficients, $\g_n$, as 
\beq
\g_{n,m}=n^{m+1}3^mC_m^{(0)}+n^{m-1}3^{m-1}C_m^{(1)}+ \cdots.
\eeq
We call $C_m^{(0)}$ spherical and $C_m^{(1)}$ torus coefficients for later convenience. 

\bthm
If we view recursive coefficients of the monic orthogonal polynomials as functions of the coupling constant $g$, 
and the size of matrix $N$, i.e. $\g_n=\g_n(g, N)$, then they admit expansions in $1/n^2$ as
\beq
\g_n(g,N)=\frac{n}{N}\left[C^{(0)}\left(\frac{3gn}{N}\right)+\frac{1}{n^2}C^{(1)}\left(\frac{3gn}{N}\right)+\cdots\right],
\eeq
where
\beq
C^{(0)}(x):=\frac{1-\sqrt{1-4x}}{2x}=\sum_{k=0}^\infty \frac{(2k)!}{(k+1)!k!} x^k,
\eeq
is the generating function of the Catalan numbers, and 
\beq
C^{(1)}(x)=\frac{2x^2}{(1-4x)^2}C^{(0)}(x)=\frac{x\left(1-\sqrt{1-4x}\right)}{(1-4x)^2}.
\eeq
\ethm
\bpf
By substitution and comparing the leading orders in $n$ of Eq.\eqref{sec_5:1}, we get, for $m\geq 0$,
\beqa
C_{m+1}^{(0)}&=&\sum_{k=0}^m C_{m-k}^{(0)}C_k^{(0)},\\
C_{m+1}^{(1)}&=&2\sum_{k=0}^m C_{m-k}^{(0)}C_k^{(1)}+\sum_{k=0}^m k(k+1)C_{m-k}^{(0)}C_k^{(0)}.\label{sec_5:2}
\eeqa

From these new sets of coupled equations, we can see that the solution of the spherical coefficients $C_m^{(0)}$ is given by the Catalan numbers,
\beq
C^{(0)}_n=\frac{(2n)!}{(n+1)!n!} \quad \Rightarrow C_0^{(0)}=1,\quad C_1^{(0)}=1, \quad C_2^{(0)}=2, \quad C_3^{(0)}=5, \cdots.
\eeq
We collect relevant information about the Catalan numbers in Appendix A.

To solve the torus coefficients, $C_m^{(1)}$, we can employ the generating function technique, which helps in transforming the coupled difference equation Eq.\eqref{sec_5:2}, into an algebraic equation 
\beq
C^{(0)}(x) := \sum_{k=0}^\infty C_k^{(0)} x^k, \quad C^{(1)}(x):=\sum_{k=0}^\infty C_k^{(1)} x^k,
\eeq
\beq\label{sec_5:3}
\Rightarrow \frac{1}{x} C^{(1)}(x)=2 C^{(0)}(x)C^{(1)}(x)+C^{(0)}(x)\frac{d}{dx}\left[x^2\frac{d}{dx}C^{(0)}(x)\right].
\eeq

Since $C^{(0)}(x)$ satisfies a quadratic equation, $x\left[C^{(0)}(x)\right]^2=C^{(0)}(x)-1$, we can also express the derivatives of $C^{(0)}(x)$ in terms of rational combinations of $C^{(0)}(x)$ (see details in Appendix A). For instance, 
\beq
\frac{d}{dx}C^{(0)}(x)=\frac{\left[C^{(0)}(x)\right]^2}{1-2xC^{(0)}(x)}=\frac{C^{(0)}(x)-1}{x\left[1-2xC^{(0)}(x)\right]}.
\eeq
Thus, we can solve $C^{(1)}(x)$ in terms of $C^{(0)}(x)$ from Eq.\eqref{sec_5:3} as
\beq\label{sec_5:3a}
C^{(1)}(x)=\frac{2x^2}{(1-4x)^2}C^{(0)}(x)=\frac{x(1-\sqrt{1-4x})}{(1-4x)^2}.
\eeq
Having solved the spherical and torus coefficients, $C_m^{(0)},C_m^{(1)}$, we can then obtain the "topological expansion" of the recursive coefficients:
\beq
\g_n=\frac{n}{N}C^{(0)}\left(\frac{3gn}{N}\right)+\frac{1}{nN}C^{(1)}\left(\frac{3gn}{N}\right)+\cdots.
\eeq
\epf
Finally, the free energy of the quartic model (up to genus one) can be computed as 
\bthm
The free energy of the quartic model $(V(x):=\frac{x^2}{2}-\frac{g}{4}x^4)$ admits a 
topological expansion in $1/N^2$ and is given by 
\beq
F_4(g,N)=N^2e_0(g)+e_1(g)+O(\frac{1}{N^2})
\eeq
with 
\beqa
e_0(g)&:=&\sum_{k=0}^\infty \frac{(2k-1)!}{k!(k+2)!}(3g)^k,\\
e_1(g)&:=&\sum_{k=0}^\infty \frac{1}{24k}\left[4^k-\frac{(2k)!}{k!k!}\right](3g)^k.
\eeqa
\ethm
\bpf
By substituting the topological expansion formula of the recursive coefficients into Eq.\eqref{sec2:2b}, we get
\beqa\label{sec4:1a}
F_4(g,N)&=&N\ln \left[h_0(g,N)\right]-N\ln\left[h_0(0,N)\right]\nonumber\\
&&+\sum_{n=1}^{N-1} (N-n) \ln\left[C^{(0)}\left(\frac{3gn}{N}\right)+\frac{1}{n^2}C^{(1)}\left(\frac{3gn}{N}\right)+O(\frac{1}{N^2})\right]\nonumber\\
&=:&N^2\left[e_0(g)+\frac{1}{N^2}e_1(g)+O(\frac{1}{N^4})\right].
\eeqa
Then using the sum of powers of the integer formula, Eq.\eqref{sec3:1a}, we get the genus zero and one contribution to the free energy of the quartic matrix model as  
\beqa
e_0(g)&:=&\sum_{k=0}^\infty \frac{(2k-1)!}{k!(k+2)!}(3g)^k,\\
e_1(g)&:=&\sum_{k=0}^\infty \frac{1}{24k}\left[4^k-\frac{(2k)!}{k!k!}\right](3g)^k.
\eeqa
These results agree with those in \cite{Bessis:1980ss}.
\epf

\section{Perturbative expansion of the free energy of the cubic model}\

The matrix potential for the cubic model is given by 
\beq
V(x):=\frac{x^2}{2}-\frac{g}{3}x^3.
\eeq
The monic orthogonal polynomials associated with the weight function $e^{-NV(x)}$ is defined through 
\beq
\int P_n(x)P_m(x)e^{-NV(x)}~dx=\delta_{mn}h_n,
\eeq
and they satisfy the famous three-term recursive relation, 
\beq
xP_n(x)=P_{n+1}(x)+\b_nP_n(x)+\a_nP_{n-1}(x).
\eeq

By using simple projection formula as discussed in Sec.$2$, we obtain a set of coupled difference equations for the recursive coefficients,
\beqa
\frac{n}{N}&=&\a_n[1-g(\b_n+\b_{n-1})],\\
\b_n&=&g(\a_{n+1}+\b_n^2+\a_n).
\eeqa

In this section, we shall perform a perturbative analysis of the the discrete difference equations to obtain a series solution of the recursive coefficients associated with the cubic model. That is, we assume that $g$ is a small number and treat both recursive coefficients $\a_n,\b_n$ as formal power series in $g$
\beqa
\a_n&=&a_{n0}+a_{n1}g+a_{n2}g^2+a_{n3}g^3+O(g^4),\\
\b_n&=&b_{n0}+b_{n1}g+b_{n2}g^2+b_{n3}g^3+O(g^4).
\eeqa
By substitution, we obtain the following solutions of the recursive coefficients:
%\beqa
%a_{n0}&=&\frac{n}{N}, \quad b_{n0}=0,\\
%a_{n1}&=&0, \quad b_{n1}=\frac{2n+1}{N},\\
%a_{n2}&=&\frac{4n^2}{N^2}, \quad b_{n2}=0,\\
%a_{n3}&=&0, \quad b_{n3}=\frac{12n^2+12n+5}{N^2},\\
%a_{n4}&=&\frac{40n^3+10n}{N^3}, \quad b_{n4}=0.
%\eeqa

\beqa
\a_n&=&\left(\frac{n}{N}\right)+\left(\frac{4n^2}{N^2}\right)g^2+\left(\frac{40n^3+10n}{N^3}\right)g^4+O(g^6),\\
\b_n&=&\left(\frac{2n+1}{N}\right)g+\left(\frac{12n^2+12n+5}{N^2}\right)g^3+O(g^5).
\eeqa
To compute the free energy, we need the following data:
\beqa
\ln \a_n(g,N)&=&\ln\left(\frac{n}{N}\right)+\left(\frac{4n}{N}\right)g^2+\left(\frac{32n^2+10}{N^2}\right)g^4+O(g^5),\\
h_0(g,N)&=&\int_{-\infty}^\infty e^{-N(\frac{x^2}{2}-\frac{g}{3}x^3)}~dx\nonumber\\
&=&\sqrt{\frac{2\pi}{N}}\left[1+\left(\frac{5}{6}\right)\frac{g^2}{N}+\left(\frac{385}{72}\right)\frac{g^4}{N^2}+O(g^6)\right].\\
\Rightarrow \ln \left[\frac{h_0(g,N)}{h_0(0,N)}\right]&=&\frac{5}{6}\left(\frac{g^2}{N}\right)+5\left(\frac{g^4}{N^2}\right)+O\left(\frac{g^6}{N^3}\right),
\eeqa
Putting all ingredients together, we have 
\beqa
F(g,N)&:=&N\ln \left[\frac{h_0(g,N)}{h_0(0,N)}\right]+\sum_{n=1}^{N-1} (N-n)\ln \left[\frac{\a_n(g,N)}{\a_n(0,N)}\right]\nonumber\\
&=&\left(\frac{2}{3}N^2+\frac{1}{6}\right)g^2+\left(\frac{8}{3}N^2+\frac{7}{3}\right)g^4+O(g^6).
\eeqa

\section{Topological expansion of the free energy of the cubic model}\

In order to derive a topological expansion (in 1/$N^2$) of the free energy of the cubic model, we first examine the structure of the recursive equations by performing a rescaling,
\beq
\ti \a_n:=N \a_n, \quad \ti \b_n:=\frac{N}{g}\b_n, \quad \e:=\frac{g^2}{N},
\eeq
\beqa
n&=&\ti\a_n[1-\e(\ti\b_n+\ti\b_{n-1})],\\
\ti\b_n&=&\ti\a_{n+1}+\ti\a_n+\e \ti\b_n^2.
\eeqa

From these expressions, we can see that:
\ben
\item[(1)] $\a_n$ and $\b_n$, as power series in $\e$, are even and odd functions in $g$, respectively,
\beqa
\a_n&=&\frac{1}{N}\sum_{k=0}^\infty a_{nk}\e^k,\quad \b_n=\frac{g}{N}\sum_{k=0}^\infty b_{nk}\e^k,\\
a_{n0}&=&0,\quad a_{nk}=\sum_{p=0}^{k-1} a_{n,k-p-1}(b_{np}+b_{n-1,p}), \quad k\geq 1,\label{sec6:1a}\\
b_{n0}&=&a_{n0}+a_{n+1,0}=2n+1,\quad b_{nk}=a_{nk}+a_{n+1,k}+\sum_{p=0}^{k-1}b_{n,k-p-1}b_{np}, \quad k\geq 1.\label{sec6:1b}
\eeqa 
\item[(2)] By mathematical induction, we can prove that the series coefficients $a_{nk},b_{nk}$, viewed as functions of $n$, must be finite polynomials of degree $k+1$. Consequently, we define 
\beqa
a_{nk}&=&n^{k+1}u^{(0)}_k+n^ku^{(1)}_k+n^{k-1}u^{(2)}_k+\cdots,\label{sec6:1}\\
b_{nk}&=&n^{k+1}v^{(0)}_k+n^kv^{(1)}_k+n^{k-1}v^{(2)}_k+\cdots.\label{sec6:2}
\eeqa
\item[(3)] Combining the two points above and substituting back to the recursive equations, Eqs.\eqref{sec6:1a}, \eqref{sec6:1b}, we get 
\beqa
u^{(0)}_k&=&2 \sum_{p=0}^{k-1} u^{(0)}_{k-p-1}v^{(0)}_p,\label{sec_7:1}\\
u^{(1)}_k&=&\sum_{p=0}^{k-1}[-(p+1)u^{(0)}_{k-p-1}v^{(0)}_p+2u^{(0)}_{k-p-1}v^{(1)}_p+2u^{(1)}_{k-p-1}v^{(0)}_p],\\
u^{(2)}_k&=&\sum_{p=0}^{k-1}[2u^{(0)}_{k-p-1}v^{(2)}_p-pu^{(0)}_{k-p-1}v^{(1)}_p+C^{p+1}_2 u^{(0)}_{k-p-1}v^{(0)}_p+2u^{(2)}_{k-p-1}v^{(0)}_p].\label{sec_7:1a}
\eeqa
\beqa
v^{(0)}_k&=&2u^{(0)}_k+\sum_{p=0}^{k-1}v^{(0)}_{k-p-1}v^{(0)}_p,\label{sec_7:3}\\
v^{(1)}_k&=&2u^{(1)}_k+(k+1)u^{(0)}_k+2\sum_{p=0}^{k-1}v^{(0)}_{k-p-1}v^{(1)}_p,\\
v^{(2)}_k&=&2u^{(2)}_k+C^{k+1}_2u^{(0)}_k+\sum_{p=0}^{k-1}[2v^{(0)}_{k-p-1}v^{(2)}_p+v^{(1)}_{k-p-1}v^{(1)}_p].\label{sec_7:2}
\eeqa
\een

\bthm
If we view the recursive coefficients as functions of the coupling constant $g$, 
and the size of matrix $N$,
\beqs
\a_n=\a_n(g,N), \quad \b_n=\b_n(g,N),
\eeqs 
then they both admit a "topological" expansion \rm{(in $1/n^2$)} as
\beq
\a_n(g,N)=\frac{n}{N}\left[u^{(0)}\left(\frac{ng^2}{N}\right)+\frac{1}{n}u^{(1)}\left(\frac{ng^2}{N}\right)+\frac{1}{n^2}u^{(2)}\left(\frac{ng^2}{N}\right)+O\left(\frac{1}{n^3}\right)\right],
\eeq
where
\beqa
u^{(0)}(x)&=&\frac{1}{24x}\{1+2\sin[\frac{2}{3} \sin^{-1}(12\sqrt{3}x)-\frac{\pi}{6}]\}\label{sec_7:4}\\
               &=&\sum_{n=0}^\infty \frac{2^{3n}x^n}{(n+1)!}\frac{\Gamma(\frac{3n+1}{2})}{\Gamma(\frac{n+1}{2})}\\
               &=&1+4x+40x^2+512x^3+7392x^4+114688x^5+\cdots,\\
u^{(1)}(x)&=&0,\\
u^{(2)}(x)&=&\frac{x [u^{(0)}]^2+72x^2u^{(0)}-x}{4(3-[u^{(0)}]^2)(1-12xu^{(0)})^3},\\
&=&\frac{-x \left\{1-72xu^{(0)}-\left[u^{(0)}\right]^2\right\}}{(\frac{15}{2}+2^5\cdot 3^4 x^2)-2^2\cdot 3^4 x u^{(0)}+(\frac{1}{2}+2^6\cdot 3^3x^2)\left[u^{(0)}\right]^2},\label{sec_7:5}
\eeqa
and
\beq
\b_n(g,N)=\frac{ng}{N}\left[v^{(0)}\left(\frac{ng^2}{N}\right)+\frac{1}{n}v^{(1)}\left(\frac{ng^2}{N}\right)+\frac{1}{n^2}v^{(2)}\left(\frac{ng^2}{N}\right)+O\left(\frac{1}{n^3}\right)\right],
\eeq
where
\beqa
v^{(0)}(x)&=&\frac{1}{2x}-\frac{1}{\sqrt{3}x}\sin[\frac{1}{3} \sin^{-1}(12\sqrt{3}x)+\frac{2\pi}{3}]\label{sec_7:6}\\
              &=&\sum_{n=0}^\infty \frac{2^{3n+1}x^n}{(n+1)!}\frac{\Gamma(\frac{3n+2}{2})}{\Gamma(\frac{n+2}{2})}\\
              &=&2+12x+128x^2+1680x^3+24576x^4+384384x^5+\cdots,\\
v^{(1)}(x)&=&\frac{1}{1-12xu^{(0)}},\\
v^{(2)}(x)&=&\frac{u^{(2)}}{2x[u^{(0)}]^2}.\label{sec_7:7}
\eeqa
\ethm
These new recursive equations can be solved by introducing the generating function technique. For instance, by defining
\beq
u^{(n)}(x):=\sum_{k=0}^\infty u^{(n)}_k x^k, \quad v^{(n)}(x):=\sum_{k=0}^\infty v^{(n)}_k x^k, \quad n=0,1,2, \cdots,
\eeq
we can transform the system of coupled difference equations Eqs.\eqref{sec_7:1}$\sim$\eqref{sec_7:2} into algebraic equations and the solutions are given as Eqs.\eqref{sec_7:4}$\sim$\eqref{sec_7:5} and Eqs.\eqref{sec_7:6}$\sim$\eqref{sec_7:7}. Please check Appendix \ref{app:b} for details.
%\beqa
%u^{(0)}(x)&=&\frac{1}{24x}\{1+2\sin[\frac{2}{3} \sin^{-1}(12\sqrt{3}x)-\frac{\pi}{6}]\}\\
%               &=&\sum_{n=0}^\infty \frac{2^{3n}}{(n+1)!}\frac{\Gamma(\frac{3n+1}{2})}{\Gamma(\frac{n+1}{2})}x^n\\
%               &=&1+4x+40x^2+512x^3+7392x^4+114688x^5+\cdots\\
%u^{(1)}(x)&=&0,\\
%u^{(2)}(x)&=&\frac{x [u^{(0)}]^2+72x^2u^{(0)}-x}{4(3-[u^{(0)}]^2)(1-12xu^{(0)})^3}.
%\eeqa
%\beqa
%v^{(0)}(x)&=&\frac{1}{2x}-\frac{1}{\sqrt{3}x}\sin[\frac{1}{3} \sin^{-1}(12\sqrt{3}x)+\frac{2\pi}{3}]\\
%              &=&\sum_{n=0}^\infty \frac{2^{3n+1}}{(n+1)!}\frac{\Gamma(\frac{3n+2}{2})}{\Gamma(\frac{n+2}{2})}x^n\\
%              &=&2+12x+128x^2+1680x^3+24576x^4+384384x^5+\cdots\\
%v^{(1)}(x)&=&\frac{1}{1-12xu^{(0)}},\\
%v^{(2)}(x)&=&\frac{u^{(2)}}{2x[u^{(0)}]^2}.
%\eeqa

Finally, the zero-th moment of the cubic model can be computed as a formal power series,
\beqa
h_0(g,N)&:=&\int e^{-NV(x)}dx\\
      &=&\sqrt{\frac{2\pi}{N}}\left[1+\frac{5}{6}\frac{g^2}{N}+\frac{385}{72}\frac{g^4}{N^2}+O\left(\frac{g^6}{N^4}\right)\right].
\eeqa
Then
\beq
\ln \left[\frac{h_0(g,N)}{h_0(0,N)}\right]=\frac{5}{6}\frac{g^2}{N}+\frac{5g^4}{N^2}+O\left(\frac{g^6}{N^4}\right).
\eeq
\bthm
The free energy of the cubic model $\left(V(x):=\frac{x^2}{2}-\frac{g}{3}x^3\right)$ admits a 
topological expansion in $1/N^2$ and is given as 
\beq
F_3(g,N)=N^2 f_0(g)+f_1(g)+O(\frac{1}{N^2})
\eeq
with 
\beq
f_0(g):=\frac{2}{3}g^2+\frac{8}{3}g^4+O\left(g^6\right),
\eeq
\beq
f_1(g):=\frac{1}{6}g^2+\frac{1}{3}g^4+O\left(g^6\right).
\eeq
\ethm
\bpf
We can now assemble all relevant equations and compute the free energy of the cubic equation as 
\begin{align} 
F(g,N)&=N\{\ln[h_0(g,N)]-\ln[h_0(0,N)]\}\tag{A}\label{aa}\\
&+\sum_{n=1}^{N-1}(N-n)\ln\left[u^{(0)}\left(\frac{g^2n}{N}\right)\right]\tag{B}\label{ab}\\
&+\sum_{n=1}^{N-1}\left(\frac{N-n}{n^2}\right)\left[\frac{u^{(2)}(\frac{g^2n}{N})}{u^{(0)}(\frac{g^2n}{N})}\right]+O(\frac{1}{N^2}).\tag{C}\label{ac}
\end{align} 
Here
\beqa
\eqref{aa}&=&\frac{5}{6}g^2+\frac{5g^4}{N}+O(\frac{1}{N^2}),\\
\eqref{ab}&=&\frac{2}{3}g^2(N^2-1)+\sum_{m=2}^\infty \frac{(\frac{3m}{2})!2^{3m}g^{2m}N^2}{3m[(m+2)!](\frac{m}{2})!}\left[1-\frac{(m+2)(m+1)}{12N^2}\right]+O(\frac{1}{N^2}),\nonumber\\
\eqref{ac}&=&\sum_{n=1}^{N-1}(N-n)\frac{g^4}{N^2}[10+408(\frac{ng^2}{N})+\cdots]\\
&=&(5-\frac{5}{N})g^4+68(1-\frac{1}{N^2})g^6+O(g^8).
\eeqa
The leading result of the cubic model free energy is given as
\beq
F(g,N)=N^2\left[\frac{2}{3}g^2+\frac{8}{3}g^4+O(g^6)\right]+N^0\left[\frac{1}{6}g^2+\frac{7}{3}g^4+O(g^6)\right]+O(\frac{1}{N^2}).
\eeq
\epf

\section{Summary and Conclusion}\

In this paper, we revisited the graphic enumeration problem. Our approach 
is based on the connections between the random matrix models and the discrete 
difference equations as derived from the orthogonal polynomial systems of 
the related matrix models.
By defining the partition function of a given matrix model (specified by a potential 
$V(M)$), 
\beq
Z(g,N):=\int \left(\prod_{k=1}^N dm_{kk}\right) \prod_{j < i} \left[d \Re m_{ij}\right]\left[d \Im_{ij}\right]e^{-N\tr V(M)}, \quad M=(m_{ij})_{i,j=1,2,\cdots, N},
\eeq
we explain that the generating functions for enumerations of random discrete 
closed surfaces consisting of regular $m$-polygons are the same as the free energies of 
the matrix model with
\beqa
V(x;g)&:=&\frac{x^2}{2}-\frac{gx^m}{m},\quad m=3,4,\\
W(\b,\g)&=&F(e^{-\b},e^\g):=\ln \left[\frac{Z(e^{-\b},e^\g)}{Z(0,e^\g)}\right]:=\sum_{n,h} e^{-n \b+(2-2h)\g}C_{n,h}.
\eeqa
Here, we treat both the partition function $Z$ and the free energy $F$ as functions 
of the coupling constant $g$ and the size of the matrix, $N$.

In addition, we make an identification between the gravitational parameters and 
the matrix model parameters as follows
\beqa
g \mbox{ (coupling constant in the matrix potential)}&=&e^{-\b} (\b: \mbox{cosmological constant}),\\
N \mbox{ (size of the finite matrix)}&=&e^{\g} (\g: \mbox{inverse Newton's constant).}
\eeqa

Our approach in this paper is based on: (1) the orthogonal polynomial representation 
of the random matrix integrals, and (2) generating function technique for solving 
difference equations.

The main results can be summarized as follows. We use two schemes to obtain series solutions of these 
difference equations. The first one is a perturbative expansion of the recursive 
coefficients in the coupling constant, $g$, and the second one is a topological 
expansion of the recursive coefficients in $1/N$. At each order (in $g$) of the perturbative expansion for the free energy of the 
matrix model, we have a finite polynomial (in $1/N^2$) whose coefficients 
correspond to the degeneracy of the random surfaces with fixed total number of 
regular polygons and specific genus. These results agree with that from the 
traditional diagrammatic expansions of the random matrix integrals. On the other hand, through a suitable reorganization of the series, we can also 
compute the topological expansion of the free energy of the matrix models. For 
genus zero (sphere) and genus one (torus), we derive explicit results of the free 
energies for both quartic and cubic models.

It is worth emphasizing that, throughout our study, we did not take any continuous 
(or integral) approximation of series summation. Hence, our computations not only 
provide independent checks of previous results \cite{DiFrancesco:1993cyw}, but also show greater applicability 
to more general cases \cite{Alexandrov:2003pj,Alexandrov:2008yq}. 
\footnote{To our knowledge, the exact finite $N$ results of the cubic model were not 
known before.}

Hopefully, our analysis of perturbative solutions to the nonlinear difference equation 
may shed some light for further explorations (e.g. asymptotic analysis, exact solutions etc.).

Based on our current results, we would like to examine the asymptotic behaviors of the 
topological expansion of the matrix model free energy. One of the immediate generalizations is to compute the correlation functions of multi-trace operators. This is of physical interest in the sense that we can check whether (1) the double-scaling limits is uniformly applicable to partition functions and the correlation functions and (2) there exist new continuous limits which correspond to various string backgrounds. In addition, it will be of great 
interest to compare with the isomonodromy analysis \cite{Fokas:1990wb, Fokas:1991za}. We believe that these might 
give useful perspective to a rigorous foundation of the double scaling limits for 
the matrix models.

\section*{Acknowledgement}

This work was initiated during a visit to the Shanghai University in 2017, where the first author 
was invited to give an introductory talk about the relation between the random matrix 
model and the discrete Painlev\'e equations. The hospitality of the host, Dajun Zhang, 
is greatly acknowledged. We would like to thank Nobutaka Nakazono for indicating the reference about the d-PII equation for the cubic matrix model, Chin-Lung Wong for the instructions of the 
connections between series solutions for algebraic equations and the Frobenius method of solving 
ordinary differential equations, Ruiming Zhang for indicating further references \cite{FlajoletSedgewick1996,FlajoletSedgewick2008} on the general 
schemes about analytic approach to the combinatoric problems, and finally the help from 
Chen-Hsun Ma for the use of computer softwares in identifying the patterns of Taylor series. 
Both authors would like 
to acknowledge the support from the National Center for Theoretical 
Sciences, and the research grant from the Ministry of Science and Technology (MOST) of Taiwan 
under the budget numbers 106-2112-M-029-003 and 104-2115-M-007-007-MY3.

\newpage

\appendix
\section{Useful Information about the Catalan Numbers}\label{app:a}\

In this appendix, we gather several useful information about the Catalan
numbers, $C_n$ \cite{wiki}, which is relevant for our study of the quartic matrix model in Sec.$3$ and Sec.$4$.

\subsection{Basic definition and explicit solution via generating function technique}\

For our purpose, it is convenient to define the Catalan numbers as a solution to the 
following nonlinear recursive relations, subject to the initial condition, $C_0=1$:
\beq\label{app-a1}
C_{n+1}=\sum_{k=0}^n C_{n-k}C_k.
\eeq
By iteration, we compute the first few terms as follows,
\beq
C_1=1,\quad C_2=2, \quad C_3=5, \quad C_4=14, \quad C_5=42, \mbox{ etc..}
\eeq

One way of obtaining the explicit formula of $C_n$ is to employ the technique of 
generating function, defined as
\beq
C(x):=\sum_{n=0}^\infty C_n x^n.
\eeq
By multiplying both sides of Eq.\eqref{app-a1} by $x^{n+1}$ and summing over $n$ 
($0\leq n<\infty$), we transform the recursive relations into an algebraic equation,
\beq\label{app-a2}
C(x)-1=x\left[C(x)\right]^2.
\eeq
One can then solve this quadratic equation and calculate the Taylor expansion of $C(x)$
around $x=0$, 
\beq\label{app-a4}
C(x)=\frac{1-(1-4x)^{\frac{1}{2}}}{2x}=\sum_{n=0}^\infty\frac{(2n)!}{(n+1)!n!} x^n.
\eeq
From this, we obtain an exact expression of the Catalan numbers as 
\beq
C_n=\frac{(2n)!}{(n+1)!n!}.
\eeq

\subsection{General approach via Frobenius method}\

One should be warned that it takes some good luck to obtain the power series solution for $C(x)$, Eq.\eqref{app-a4}, from 
the algebraic equation, Eq.\eqref{app-a2}, due to the simplicity of the structure. 

In general, a complete solution of a given algebraic equation in terms of power series is never a trivial task 
(this will become evident when we move on to the case of the cubic matrix model). For this reason, we shall illustrate 
the Frobenius method in solving the power series form of $C(x)$ as follows.

It proves to be very useful if we take derivative (w.r.t $x$) on Eq.\eqref{app-a2}, and after rearranging
\beq\label{app-a3}
C'(x)=\frac{\left[C(x)\right]^2}{1-2xC(x)}=\frac{-1+C(x)}{x\left[1-2xC(x)\right]}.
\eeq

From the final equality of Eq.\eqref{app-a3}, it is natural to expect that $C'(x), C(x)$ and $1$ are algebraically 
dependent. That is, there exist polynomials $\a(x),\b(x)$ such that 
\beq\label{app-a5}
\a C'+\b C+1=0.
\eeq
Indeed, by substitution (using Eq.\eqref{app-a3}) and factoring out the common denominator, we have,
\beq
(\a-\b x-2x^2)C(x)=\a-2\b x-x.
\eeq
Demanding both coefficient functions to be zero, we get
\beq
\a(x)=4x^2-x, \mbox{ and } \b(x)=2x-1.
\eeq
In this way, we transform the algebraic equation, Eq.\eqref{app-a2}, into a inhomogeneous ordinary 
differential equation, Eq.\eqref{app-a5}. Then the standard Frobenius method can be applied, and 
we derive a linear recursive relation among the Taylor coefficients of $C(x)$ from Eq.\eqref{app-a5} as
\beq
C(x)=\sum_{n=0}^\infty C_n x^n \Rightarrow C_n=\frac{2(2n-1)}{n+1}C_{n-1}.
\eeq
Finally, by telescoping, we obtain the exact expressions of the Catalan number
\beq
C_n=\frac{(2n)!}{(n+1)!n!}.
\eeq
\subsection{Higher derivatives in terms of the rational expressions}\

From the first equality of Eq.\eqref{app-a3}, we get
\beqa
\frac{d}{dx}\left[\ln C(x)\right]&=&\frac{C'(x)}{C(x)}=\frac{C(x)}{1-2xC(x)}\\
&=&\left[\frac{1}{1-2xC(x)}-1\right]\frac{1}{2x}=\left[(1-4x)^{-\frac{1}{2}}-1\right]\frac{1}{2x}\\
&=&\sum_{k=0}^\infty \frac{(2k+1)!}{(k+1)!k!} x^k.
\eeqa
Hence, 
\beq
\ln C(x)=\sum_{k=1}^\infty \frac{(2k-1)!}{k!k!} x^k.
\eeq
This result is used in Eq.\eqref{sec4:1a} for the topological expansion of the free energy of the quartic matrix model. The second equality of Eq.\eqref{app-a3} implies that we can always express any 
derivative of $C(x)$, $\frac{d}{dx} C(x)$, as a rational combination of $C(x)$ in the 
first degree. Indeed, we have 
\beqa
C'(x)&=&\frac{\left[C(x)\right]^2}{1-2xC(x)}=\frac{-1+C(x)}{x-2x^2C(x)},\\
C''(x)&=&\frac{(2-6x)+(-2+8x)C(x)}{x^2(1-4x)\left[1-2xC(x)\right]},\\
\mbox{and }\left[x^2C(x)\right]'&=&x^2C''(x)+2xC'(x)=\frac{2x}{(1-4x)\left[1-2xC(x)\right]}.
\eeqa
The last equation is used in deriving genus-one contribution to the free energy of 
the quartic matrix model, Eqs.\eqref{sec_5:3} $\sim$ \eqref{sec_5:3a}.

\newpage
\section{Useful information for solving the cubic model}\label{app:b}\
In the previous appendix, we invoked the generating function technique to solve a non-linear 
recursive relation, Eq.\eqref{app-a1}. Here, we apply the same idea to embrace three types of equations,
 and solve for the recursive coefficients associated with the cubic matrix model.

\begin{figure}[h]
       \begin{center}
       \includegraphics[scale=0.4]{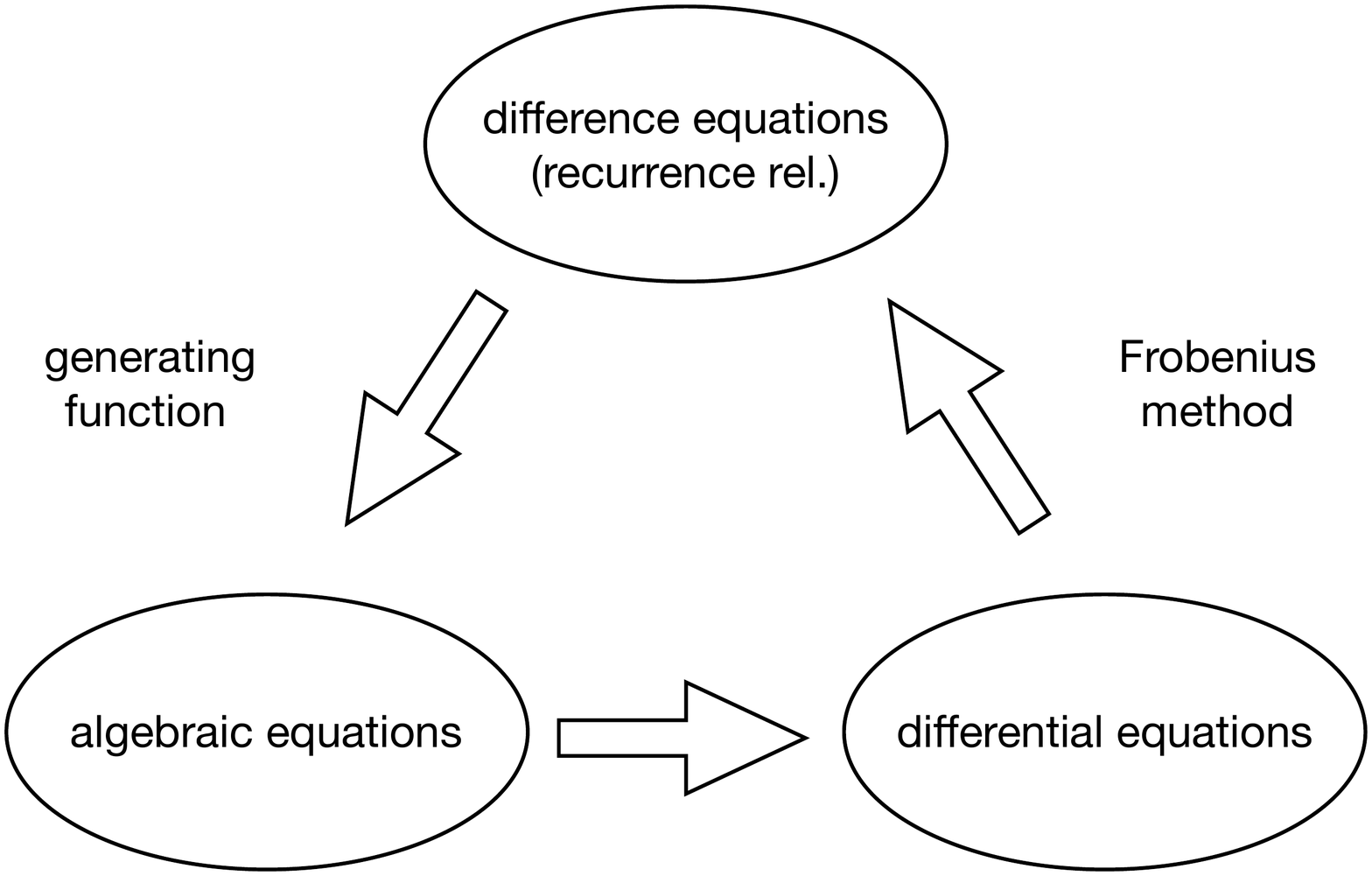}
       \captionof{figure}{Interconnections among three types of equations}\label{fig:3}
       \end{center}
\end{figure}

\subsection{Solutions of the generating functions for the spherical coefficients $u_m^{(0)},v_m^{(0)}$}\

By defining the generating functions of the spherical coefficients, $u^{(0)}_m, v^{(0)}_m$, Eqs.\eqref{sec6:1}, \eqref{sec6:2}
\beq
u^{(0)}(x):=\sum_{m=0}^\infty u^{(0)}_m x^m, \quad v^{(0)}(x):=\sum_{m=0}^\infty v^{(0)}_m x^m,
\eeq
we transform the infinite sets of coupled difference equations, Eqs.\eqref{sec_7:1},\eqref{sec_7:3}, into two coupled algebraic equations
\beqa
u^{(0)}(x)-u^{(0)}_0&=&2xu^{(0)}(x)v^{(0)}(x),\\
v^{(0)}(x)-v^{(0)}_0&=&2u^{(0)}(x)-2u^{(0)}_0+x\left[v^{(0)}(x)\right]^2.
\eeqa
Note that $u^{(0)}_0=1$ and $v^{(0)}_0=2$. Upon substitution, we get
\beq
u^{(0)}(x)=\frac{1}{1-2xv^{(0)}(x)},\quad
v^{(0)}(x)=\frac{u^{(0)}(x)-1}{2x u^{(0)}(x)}.
\eeq
Further substitutions show that $u^{(0)}(x)$ and $v^{(0)}(x)$ are solutions to the following cubic equations:
\beqa
8x\left[u^{(0)}(x)\right]^3-\left[u^{(0)}(x)\right]^2+1=0,\\
2x^2\left[v^{(0)}(x)\right]^3-3x\left[v^{(0)}(x)\right]^2+v^{(0)}(x)-2=0.\label{app_b1}
\eeqa
Using the triple-angle formula, the solutions satisfying the "initial conditions", 
\beq
\lim_{x \rightarrow 0} u^{(0)}(x)=u^{(0)}_0=1, \mbox{ and } \lim_{x\rightarrow 0} v^{(0)}(x)=v^{(0)}_0=2,
\eeq
are given as 
\beqa
u^{(0)}(x)&=&\frac{1}{24x}\left\{1+2\sin\left[\frac{2}{3} \sin^{-1}(12\sqrt{3}x)-\frac{\pi}{6}\right]\right\},\\
v^{(0)}(x)&=&\frac{1}{2x}-\frac{1}{\sqrt{3}x}\sin\left[\frac{1}{3} \sin^{-1}(12\sqrt{3}x)+\frac{2\pi}{3}\right].
\eeqa

\subsection{Taylor expansions of the spherical generating functions}\

While there are standard results of the Taylor expansion for the (inverse) trigonometric functions, the compositions of power series expansions into a  single Taylor expansion is a totally nontrivial task. For this reason, we need to adjust the forms of cubic equations Eq.\eqref{app_b1} by defining
\beq
\tau:=12\sqrt{3}x, \quad w(\tau):=\frac{\sqrt{3}}{2}-\frac{\tau}{12}v^{(0)}(x).
\eeq
One can check that, after the change of variables, we get
\beq
4w^3-3w+\tau=0.
\eeq
Consulting the formula in \cite{WhittakerWatson1927}, we obtain the Taylor expansion of $v^{(0)}(x)$ as
\beqa
v^{(0)}(x)&=& \sum_{n=0}^\infty \frac{2^{3n+1}x^n}{(n+1)!}\frac{\Gamma{\left(\frac{3n+2}{2}\right)}}{\Gamma{\left(\frac{n+2}{2}\right)}}\\
&=&2+12x+128x^2+1680x^3+24576x^4+384384x^5+\cdots .
\eeqa

To derive the series expansion of $u^{(0)}(x)$ around $x=0$, we use computer software to calculate the first $10$ terms in the Taylor expansion of $u^{(0)}(x)$. By carefully studying the pattern of the Taylor coefficients (prime factor decomposition), we deduce the general formula as follows.
\beqa
u^{(0)}(x)&=& \sum_{n=0}^\infty \frac{2^{3n}x^n}{(n+1)!}\frac{\Gamma{\left(\frac{3n+1}{2}\right)}}{\Gamma{\left(\frac{n+1}{2}\right)}}\\
&=&1+4x+40x^2+512x^3+7392x^4+114688x^5+\cdots .
\eeqa

To compute the free energy of the cubic matrix model, we also need the following results, which can be obtained in a similar vein,
\beq
\ln \left[u^{(0)}(x)\right]=\sum_{n=1}^\infty \frac{\left(\frac{3n}{2}\right)!}{(3n)n!\left(\frac{n}{2}\right)!}(8x)^n.
\eeq

\subsection{Derivatives of higher genus generating functions in terms of the spherical generating functions}\

As emphasized in the Appendix A, since the spherical generating functions $u^{(0)}(x),v^{(0)}(x)$ satisfy cubic algebraic equations, we can always express any higher derivatives of them in terms of rational combinations of $u^{(0)}(x), v^{(0)}(x)$ of degree $2$. We can then feedback these rational combinations into the coupled sets of equations, Eqs.\eqref{sec_7:1}$\sim$\eqref{sec_7:1a},\eqref{sec_7:3}$\sim$\eqref{sec_7:2}, to solve for the higher-genus generating functions. Since the computations are straightforward, we simply collect the relevant results for reference.
\beqa
\frac{d}{dx}u^{(0)}(x)&=&\frac{4\left[u^{(0)}(x)\right]^2}{1-12x u^{(0)}(x)},\\
\frac{d}{dx}v^{(0)}(x)&=&\frac{-3x\left[v^{(0)}(x)\right]^2+2v^{(0)}(x)-4}{6x^3\left[v^{(0)}(x)\right]^2-6x^2v^{(0)}(x)+x}\\
&=&\frac{12x\left[u^{(0)}(x)\right]^2-(1+8x)u^{(0)}(x)+1}{-24x^3\left[u^{(0)}(x)\right]^2+2x^2u^{(0)}(x)},\\
\frac{d}{dx}\left[xu^{(0)}(x)\right]&=&\frac{1}{u^{(0)}\left[1-12x u^{(0)}(x)\right]},\\
\frac{d}{dx}\left[xv^{(0)}(x)\right]&=&\frac{2}{1-12x u^{(0)}(x)},\\
u^{(1)}(x)&=&0 \quad \Rightarrow \frac{d}{dx}u^{(1)}(x)=0,\\
 \frac{d}{dx}v^{(1)}(x)&=&\frac{12}{u^{(0)}\left[1-12x u^{(0)}\right]^3}.
\eeqa

\bibliographystyle{plain}
%\bibliography{TeXBiB}

\begin{thebibliography}{99}
%\cite{Bessis:1980ss}
\bibitem{Bessis:1980ss} 
  D.~Bessis, C.~Itzykson and J.~B.~Zuber,
  Quantum field theory techniques in graphical enumeration,
  Adv.\ Appl.\ Math.\  {\bf 1}, 109 (1980).
  
%\cite{Bessis:1979is}
\bibitem{Bessis:1979is} 
  D.~Bessis,
  A New Method in the Combinatorics of the Topological Expansion,
  Commun.\ Math.\ Phys.\  {\bf 69}, 147 (1979).
  
%\cite{DiFrancesco:1993cyw}
\bibitem{DiFrancesco:1993cyw} 
  P.~Di Francesco, P.~H.~Ginsparg and J.~Zinn-Justin,
  2-D Gravity and random matrices,
  Phys.\ Rept.\  {\bf 254}, 1 (1995).
  %%CITATION = doi:10.1016/0370-1573(94)00084-G;%%
  %660 citations counted in INSPIRE as of 03 Nov 2017

%\cite{Gross:1989vs}
\bibitem{Gross:1989vs} 
  D.~J.~Gross and A.~A.~Migdal,
  Nonperturbative Two-Dimensional Quantum Gravity,
  Phys.\ Rev.\ Lett.\  {\bf 64}, 127 (1990).

  %%CITATION = doi:10.1103/PhysRevLett.64.127;%%
  %953 citations counted in INSPIRE as of 02 Nov 2017
  
%\cite{Gross:1989aw}
\bibitem{Gross:1989aw} 
  D.~J.~Gross and A.~A.~Migdal,
  A Nonperturbative Treatment of Two-dimensional Quantum Gravity,
  Nucl.\ Phys.\ B {\bf 340}, 333 (1990).

  %%CITATION = doi:10.1016/0196-8858(80)90008-1;%%
  %287 citations counted in INSPIRE as of 02 Nov 2017
%\cite{Brezin:1990rb}
\bibitem{Brezin:1990rb} 
  E.~Brezin and V.~A.~Kazakov,
  Exactly Solvable Field Theories of Closed Strings,
  Phys.\ Lett.\ B {\bf 236}, 144 (1990).
  
 %\cite{Douglas:1989ve}
\bibitem{Douglas:1989ve} 
  M.~R.~Douglas and S.~H.~Shenker,
  Strings in Less Than One-Dimension,
  Nucl.\ Phys.\ B {\bf 335}, 635 (1990).
  
\bibitem{doCarmo:1976}
Manfredo P.~do Carmo, Differential Geometry of Curves and Surfaces, Prentice-Hall, (1976).
 
\bibitem{M60}
M. L. Mehta, On the statistical properties of the level-spacings in nuclear spectra, Nuclear Phys. v. 18 (1960) 395-419.

\bibitem{MG60}
M. L. Mehta; M. Gaudin, On the density of eigenvalues of a random matrix, Nuclear Phys. 18 (1960) 420-427. 
 
%\cite{tHooft:1973alw}
\bibitem{tHooft:1973alw} 
  G.~'t Hooft,
  A Planar Diagram Theory for Strong Interactions,
  Nucl.\ Phys.\ B {\bf 72}, 461 (1974).
    
%\cite{Brezin:1977sv}
\bibitem{Brezin:1977sv} 
  E.~Brezin, C.~Itzykson, G.~Parisi and J.~B.~Zuber,
  Planar Diagrams,
  Commun.\ Math.\ Phys.\  {\bf 59}, 35 (1978).

  %%CITATION = doi:10.1007/BF01614153;%%
  %1205 citations counted in INSPIRE as of 02 Nov 2017
  


  %%CITATION = doi:10.1007/BF01221445;%%
  %104 citations counted in INSPIRE as of 02 Nov 2017
\bibitem{KNY2017}
K.~Kajiwara, M.~Noumi and Y.~Yamada, 
Geometric Aspects of Painlev\'e Equations,
J.\ Phys.\ A: Math. and Theo., Vol. 50, Num. 7 (2017).

%\cite{Fokas:1990wb}
\bibitem{Fokas:1990wb} 
  A.~S.~Fokas, A.~R.~Its and A.~V.~Kitaev,
  Discrete Painleve equations and their appearance in quantum gravity,
  Commun.\ Math.\ Phys.\  {\bf 142}, 313 (1991).

\bibitem{BA96}
A. Beardon, Sums of Powers of Integers, Amer. Math. Monthly 103 (1996), 201-213.

%\cite{Alexandrov:2003pj}
\bibitem{Alexandrov:2003pj} 
  A.~S.~Alexandrov, A.~Mironov and A.~Morozov,
  Partition Functions of Matrix Models as the First Special Functions of String Theory. 1. Finite size Hermitean one matrix model,
  Int.\ J.\ Mod.\ Phys.\ A {\bf 19}, 4127 (2004).
  %%CITATION = doi:10.1142/S0217751X04018245;%%
  %109 citations counted in INSPIRE as of 03 Nov 2017

%\cite{Alexandrov:2008yq}
\bibitem{Alexandrov:2008yq} 
  A.~S.~Alexandrov, A.~Mironov, A.~Morozov and P.~Putrov,
  Partition Functions of Matrix Models as the First Special Functions of String Theory. II. Kontsevich Model,
  Int.\ J.\ Mod.\ Phys.\ A {\bf 24}, 4939 (2009).
  %%CITATION = doi:10.1142/S0217751X09046278;%%
  %55 citations counted in INSPIRE as of 03 Nov 2017



  %%CITATION = doi:10.1016/0550-3213(74)90154-0;%%
  %4381 citations counted in INSPIRE as of 03 Nov 2017
  


  %%CITATION = doi:10.1016/0550-3213(90)90450-R;%%
  %427 citations counted in INSPIRE as of 02 Nov 2017



  %%CITATION = doi:10.1016/0370-2693(90)90818-Q;%%
  %1119 citations counted in INSPIRE as of 02 Nov 2017



  %%CITATION = doi:10.1016/0550-3213(90)90522-F;%%
  %1028 citations counted in INSPIRE as of 02 Nov 2017



  %%CITATION = doi:10.1007/BF02102066;%%
  %30 citations counted in INSPIRE as of 02 Nov 2017

%\cite{Fokas:1991za}
\bibitem{Fokas:1991za} 
  A.~S.~Fokas, A.~R.~Its and A.~V.~Kitaev,
  The Isomonodromy approach to matrix models in 2-D quantum gravity,
  Commun.\ Math.\ Phys.\  {\bf 147}, 395 (1992).

  %%CITATION = doi:10.1007/BF02096594;%%
  %53 citations counted in INSPIRE as of 02 Nov 2017

\bibitem{FlajoletSedgewick1996}
P. Flajolet, R. Sedgewick, An Introduction to the Analysis of Algorithms, Addison Wesley, 1996.

\bibitem{FlajoletSedgewick2008}
P. Flajolet, R. Sedgewick, Analytic Combinatorics, Cambridge University Press, 2008.
\bibitem{wiki}
https://en.wikipedia.org/wiki/Catalan$\underline{\hspace{0.5em}}$number

\bibitem{WhittakerWatson1927}
E. T. Whittaker and G. N. Watson, A Course of Modern Analysis, Cambridge University Press, 1927.




\end{thebibliography}

\end{document}